\documentclass[pdflatex,sn-mathphys-num]{sn-jnl}

\usepackage{graphicx}%
\usepackage{multirow}%
\usepackage{amsmath,amssymb,amsfonts}%
\usepackage{amsthm}%
\usepackage{mathrsfs}%
\usepackage[title]{appendix}%
\usepackage{xcolor}%
\usepackage{textcomp}%
\usepackage{manyfoot}%
\usepackage{booktabs}%
\usepackage{algorithm}%
\usepackage{algorithmicx}%
\usepackage{algpseudocode}%
\usepackage{listings}%
\usepackage{ulem}
\usepackage{comment}

\def \bv {\mathbf{v}}
\def \HH {\mathcal{H}}
\def \bn {\mathbf{n}}
\def \epH {\epsilon_\mathcal{H}}
\def \hp {\hat{p}}

\theoremstyle{thmstyleone}%

\theoremstyle{thmstyletwo}%

\theoremstyle{thmstylethree}%

\begin{document}

\title[Testing the Equivalence Principle in Galaxy Clusters]{Testing the Equivalence Principle in Galaxy Clusters}

\author[1]{\fnm{Enea} \sur{Di Dio}}

\author[2,1]{\fnm{Sveva} \sur{Castello}}

\author[1]{\fnm{Camille} \sur{Bonvin}}

\affil[1]{Universit\'e de Gen\`eve, D\'epartement de Physique Th\'eorique and Centre for Astroparticle Physics,
24 quai Ernest-Ansermet, CH-1211 Gen\`eve 4, Switzerland}

\affil[2]{Max-Planck-Institut f\"ur Astrophysik, Karl-Schwarzschild-Str.~1, 85748 Garching, Germany}

\abstract{Clusters of galaxies have been used to measure a subtle effect predicted by Einstein: gravitational redshift. This signal encodes pristine information about our Universe, since it is sensitive to the depth of the clusters' gravitational potential wells. In this work, we show how gravitational redshift can be used to test a fundamental physical principle: the weak equivalence principle. This principle stipulates that all matter falls in the same way in a gravitational potential, regardless of its nature. By comparing the amplitude of the gravitational redshift signal with the velocity dispersion in galaxy clusters, we build a novel test of this principle targeted to the unknown dark matter. Our test is sensitive to any additional interaction that would alter the way dark matter falls in gravitational potentials, hence leading to a violation of the equivalence principle. We show that currently available data can constrain the presence of a fifth force in clusters at the level of 7-14\%, while the newest surveys will reach a precision of a few percents. This demonstrates the crucial role played by galaxy clusters in testing fundamental properties of dark matter.}

\maketitle

\begingroup
\normalsize
\noindent
Emails: \texttt{enea.didio@unige.ch},
\texttt{svevacas@mpa-garching.mpg.de},
\texttt{camille.bonvin@unige.ch}
\par
\endgroup

\section{Introduction}\label{sec:intro}

One of the key open questions in fundamental physics concerns the nature of the unknown dark matter. Despite the overwhelming evidence of its existence, it has never been detected, and its fundamental properties remain largely unknown. As the largest gravitationally bound objects in the Universe, galaxy clusters represent the ideal cosmic laboratory to probe the behaviour of dark matter and determine whether it is consistent with the simplest Cold Dark Matter (CDM) paradigm, i.e.~cold, collisionless, non-interacting particles.

Various methods have been used to infer dark matter properties from galaxy clusters. A standard approach consists in measuring the distribution of visible and dark matter from gravitational lensing, thus reconstructing the matter density in clusters. The reconstructed profiles can then be used to look for signatures of models beyond the CDM paradigm, which modify the density profile and shape of clusters. For example, self-interacting dark matter lowers the central density and produces rounder cluster cores~\cite{Spergel:1999mh,Brinckmann:2017uve,Robertson:2018anx}, while warm dark matter models instead tend to decrease the number of substructures~\cite{Bode:2000gq,Elahi:2014gwa}. Another useful test consists in using velocity dispersions in clusters or X-ray emissions to reconstruct the cluster mass and compare it with the one inferred from gravitational lensing, in order to identify any mismatch that could indicate non-standard dark matter, see e.g.~\cite{Sagunski:2020spe,Adhikari:2022sbh,Harvey:2024gpk}.

In this work, we propose a new test to probe a fundamental property of dark matter: whether it obeys or not the weak equivalence principle.  According to this principle, all matter in the Universe falls in the same way under gravity, independently of its composition or structure. This has been tested with exquisite precision in laboratories for all types of ordinary matter (see e.g.~\cite{Wagner:2012ui}), while its validity for dark matter remains a crucial open question. In recent years, various methods have been proposed to answer this question, using multiple observables from cosmological scales~\cite{Creminelli:2013nua,Kehagias:2013rpa,Bonvin:2018ckp,Bonvin:2022tii,Castello:2024jmq,Castello:2024lhl,Dam:2025kmr} down to galactic scales~\cite{Desmond:2018kdn,Desmond:2018sdy,Pardo:2019wie}, as well as with astrophysical sources~\cite{Bartlett:2021yyp, Reischke:2023gjv}. In this paper, we develop a method to test the weak equivalence principle using galaxy clusters. The key idea is to compare the depth of the clusters' gravitational potentials with the velocity of the member galaxies, which are mostly constituted of dark matter. 
If dark matter obeys the weak equivalence principle, these two quantities are related through the Jeans equation, which describes the motion of the galaxies under the effect of gravity. In contrast, if the weak equivalence principle is broken, galaxies can fall faster or slower under the influence of an additional, non-gravitational interaction. Crucially, this test requires measuring the {\it time} distortion of the geometry, which is the quantity that directly governs the motion of galaxies through the Jeans equation. 

This test has never been performed so far. Its importance has been demonstrated on linear cosmological scales~\cite{Bonvin:2018ckp,Bonvin:2022tii,Castello:2024jmq,Castello:2024lhl}, where however measurements of the time distortion require very large numbers of galaxies and have not been possible so far. Here, we demonstrate that galaxy clusters provide the ideal environment to perform this test. The crucial feature of clusters is that both quantities involved in the test, the velocity of galaxies and the time distortion, can be measured from a single signal: the redshift difference between the clusters members and the Bright Central Galaxy (BCG), $\Delta z\equiv z_{\rm member}-z_{\rm BCG}$. Both the velocity of galaxies and the time distortion indeed affect $\Delta z$: the former through Doppler effects and the latter through gravitational redshift, which corresponds to a frequency change of photons due to the time distortion in the cluster's gravitational potential. Note that gravitational lensing, in contrast to gravitational redshift, does not measure directly the time distortion, but rather the Weyl potential, i.e.~the sum of the distortion of time and that of space.

The impact of the time distortion and of the galaxy velocities on $\Delta z$ can be separated thanks to their different symmetries. The difference in velocities between the cluster members and the BCG can take any sign, generating a spread in $\Delta z$. In contrast, the time distortion always generates a negative difference $\Delta z<0$, since the BCG lies at the bottom of the gravitational potential and is therefore always more redshifted than the cluster members. These symmetries can be exploited by constructing a histogram for $\Delta z$ in a sample of clusters. The width of the distribution provides a direct measurement of the galaxy velocities, while the negative shift of the mean provides a measurement of the time distortion. This method has been successfully used to measure gravitational redshift and thus the time distortion in galaxy clusters from various surveys~\cite{Wojtak:2011ia,Sadeh:2014rya,Rosselli:2022qoz,eBOSS:2021ofn}.

In the following, we present a method to directly test the validity of the weak equivalence principle by combining the width and the shift of the redshift  distribution. We show how to account for second-order Doppler effects that act as a contaminant to gravitational redshift and hence need to be consistently included in the test in order not to bias the result, as shown in \cite{DiDio:2025bff}. We demonstrate that, by construction, our test is sensitive to theories that break the weak equivalence principle for dark matter and not to alternative theories of gravity that modify the way all matter falls under gravity. Such theories modify the depth of gravitational potentials and the velocity of galaxies by changing the way matter accretes in the Universe, but they do not affect the link between galaxy velocities and time distortions, so that the weak equivalence principle is still valid.\footnote{A modified theory of gravity that couples differently to dark matter and ordinary matter would also break the weak equivalence principle, see e.g.~\cite{Blas:2012vn,Gleyzes:2015pma}. However, from an observational point of view, this is indistinguishable from a modified theory of gravity that universally couples to all matter {\it plus} a non-gravitational force acting on dark matter.  Hence, a logical way of drawing the line between modified theories of gravity and non-gravitational dark matter interactions is to call modified gravity the theories that couple universally to all matter, respecting the weak equivalence principle, see e.g.~\cite{Joyce:2016vqv}.} In contrast, scenarios where dark matter is affected by a fifth non-gravitational force, or is coupled to dark energy or to dark radiation would break the weak equivalence principle and are therefore detectable with our test. 

We forecast the precision of our test with current and future data.  With $10^5$ galaxies, as in the first detection of gravitational redshift in clusters \cite{Wojtak:2011ia}, the test is sensitive to deviations in the weak equivalence principle at the level of 7-14\%,  depending on the prior knowledge on astrophysical parameters. By providing an increased number of observed galaxies, the newest generation of surveys like the Dark Energy Spectroscopic Instrument\footnote{\url{https://www.desi.lbl.gov/}} (DESI) and the Euclid satellite\footnote{\url{https://www.esa.int/Science_Exploration/Space_Science/Euclid}} will tighten the constraints down to a few percent.

\section{The Jeans equation and the weak equivalence principle}
\label{sec:WEP}

If dark matter only interacts gravitationally, its energy and momentum are conserved. This is no longer the case in the presence of non-gravitational interactions, through which dark matter can exchange energy and/or momentum with other constituents. In Appendix~\ref{app:distribution}, we derive the impact of such non-gravitational interactions on the dynamics of a galaxy cluster in full generality. We start from the Vlasov equation, which governs the evolution of the galaxy distribution function in clusters, and show how dark matter interactions generically modify the geodesic equation, directly impacting the motion of galaxies. This leads to a finite number of additional terms in the continuity and Euler equations, derived in a model-independent way. In turn, these modify the Jeans equation, which relates the cluster velocity dispersion to the time distortion $\Psi$. Restricting for simplicity to scenarios  where dark matter exchanges only momentum, the Jeans equation in the presence of non-gravitational interactions takes the form
\begin{equation}
\label{eq:Jeans}
    \partial_Z \left( \rho \sigma^2_v \right) + \left( 1+ \Gamma \right) \rho \, \partial_Z \Psi =0  \, .
\end{equation}
Here, $\rho$ is the density field, $\sigma_v^2$ the one-dimensional velocity dispersion of galaxies and $\partial_Z$ is the derivative along the cylindrical coordinate $Z$. The parameter $\Gamma$ is the key quantity of our analysis, as it encodes the amplitude of a potential fifth force acting on dark matter. 

From Eq.~\eqref{eq:Jeans}, it becomes clear that by comparing the velocity dispersion, $\sigma_v^2$, measured from the width of the redshift difference in clusters, with the time distortion $\Psi$, measured from the shift of the mean, we can directly test for the presence of a non-zero fifth force, i.e.~$\Gamma\neq 0$. Crucially, modified theories of gravity do not lead to $\Gamma\neq 0$. As such, testing the Jeans equation by comparing the velocity dispersion with the time distortion of the metric $\Psi$ provides a way of testing the behaviour of dark matter {\it independently} of the underlying theory of gravity (the Einstein equations were never used in our derivation in Appendix \ref{app:distribution}.). Our test is therefore orthogonal to previous tests in clusters that combine measurements of the clusters dynamics and of gravitational lensing, see e.g.~\cite{Schmidt:2010jr,Jain:2010ka,Lam:2012by}. Such tests compare the gravitational lensing signal with the mass inferred from velocity dispersion or X-rays observations, so that they are directly sensitive to modifications of gravity. In particular, they probe the link between the sum of the gravitational potentials, $\Phi+\Psi$, and the masses of the clusters, which is generically modified in gravity theories beyond general relativity. By contrast, by using measurements of $\Psi$ and directly probing for the presence of $\Gamma$ in the Jeans equation, our test uniquely targets a breaking of the weak equivalence principle for dark matter, which cannot be generated by modifications of gravity. 

\section{Observing galaxy velocities and gravitational redshift in clusters}

We now show how to relate the key quantities entering the Jeans equation in Eq.~\eqref{eq:Jeans} to the observables that we can measure in clusters, namely the shift and width of the redshift difference. As explained in Section \ref{sec:intro}, the first step in the measurement consists in taking the redshift difference between each cluster member and the BCG. The distribution is then constructed by combining all BCG-member pairs separated by a fixed transverse direction $R_\perp$, in order to probe the depth of the gravitational potential at different distances from the cluster centre. Since we expect to observe around $10$ members per cluster, we need to stack several clusters in order to properly characterise the redshift distribution and extract its mean and width.

In~\cite{DiDio:2025bff}, we derived a model for the mean of this signal based on the so-called weak-field expansion, which is the relevant perturbation scheme valid on the scales of clusters. The final result reads
\begin{align}
\label{eq:mean}
\langle \Delta z \rangle_{R_\perp} = -  \langle \Delta \Psi \rangle_{R_\perp}  + \left( \frac{3}{2} - \mathcal{R}\right) \sigma_{\rm los}^2(R_\perp)  - \frac{7}{2} \sigma_{\rm BCG}^2\, ,    
\end{align}
where the brackets $\langle \rangle$ denote a theoretical average over the velocity distribution and over all stacked clusters.\footnote{In~\cite{DiDio:2025bff}, we also introduced the superscript ``stacked'' to distinguish the average over a single cluster from the one over stacked clusters. For simplicity, we drop this superscript here, since we will always consider the average over stacked clusters.} As expected, the mean of the redshift difference at fixed $R_\perp$ does not vanish. Thus, the distribution is not centred on zero but is instead shifted away from it due to a combination of different effects. The first one is the signal we are after, i.e.~gravitational redshift, averaged over all stacked clusters: 
\begin{align} 
-  \langle \Delta \Psi \rangle_{R_\perp}= \Sigma^{-1}\int dM \frac{dN_c}{dM} \int dr_e \ \rho_g^{\rm real} \left(r_e, F\ge F_* \right) \left(\Psi_{\rm BCG}-\Psi(r_e) \right)\,  .
\end{align}
The first integral is taken over the mass of the clusters $M$ and represents the average over the stacked clusters, weighted by the cluster mass function $dN_c/dM$. The integral over the coordinate $r_e$ encodes the average over all cluster members along the line-of-sight at fixed transverse separation $R_\perp$ from the BCG. The subscript $e$ indicates that the average is performed at fixed time in the rest frame of the BCG (see~\cite{DiDio:2025bff} for more detail). The gravitational redshift difference between the BCG and the cluster member, $\Psi_{\rm BCG}-\Psi(r_e)$, is negative and is weighted by the density of the galaxies along the line-of-sight whose flux is larger than the observational threshold $F_*$. The superscript ``real'' means that this density corresponds to the one in real space, i.e.~expressed as a function of position, which we can model through a known functional form such as the Navarro-Frenk-White (NFW) one~\cite{Navarro:1995iw}. Lastly, $\Sigma$ is a normalisation over the density of clusters, defined as 
\begin{equation}
\label{eq:sigma_R}
    \Sigma(R_\perp)\equiv\int dM \frac{dN_c}{dM}\int dr_e \ \rho_g^{\rm real} \left(r_e, F\ge F_* \right) \, .
\end{equation}

As can be noticed from Eq.~\eqref{eq:mean}, gravitational redshift is not the only effect that shifts the position of the mean away from zero. Second-order Doppler terms from the cluster members and from the BCG also contribute, since they always have the same sign. By contrast, the linear Doppler shift can be either positive or negative and hence vanishes when integrating over the velocity distribution function of the clusters. As in the case of gravitational redshift, the contribution from the line-of-sight velocity dispersion is averaged over all masses $M$ and all distances $r_e$,
\begin{align}
\sigma_{\rm los}^2(R_\perp)= \Sigma^{-1}\int dM \frac{dN_c}{dM} \int dr_e \ \rho_g^{\rm real} \left(r_e, F\ge F_* \right) \sigma_v^2(r_e)\,  .
\end{align}
Within the weak field expansion, $\sigma_{\rm los}^2$ and $\Delta\Psi$ are of the same order of magnitude in clusters and, as first pointed out in~\cite{Zhao:2012gxk}, neglecting second-order Doppler effects is not consistent. Eq.~\eqref{eq:mean} consistently includes all Doppler contributions to the shift that are of the same order as gravitational redshift.\footnote{In~\cite{DiDio:2025bff}, we defined a weak-field parameter $\epsilon_\HH=\HH/k$, where $\HH$ is the Hubble parameter and $k$ the wavenumber. This quantity is significantly smaller than 1 on cluster scales and can thus be used as the reference quantity in the expansion. In~\cite{DiDio:2025bff}, we consistently computed the shift including all effects up to order $\mathcal{O}(\epsilon^2_\HH)$.} The term proportional to $\sigma_{\rm los}^2$ appears with the prefactor $\frac{3}{2} - \mathcal{R}=\frac{3}{2} - \frac{5}{2} s_b \left( 3 + \alpha \right)$, which encodes all Doppler contributions, from the second-order transverse Doppler effect but also from the product of linear Doppler magnification and linear Doppler effects. This expression contains two free parameters that have to be measured for the galaxy population considered: the magnification bias $s_b$ and the spectral index $\alpha$, see \cite{DiDio:2025bff} for a detailed discussion. 

The last contribution to the shift of the mean in Eq.~\eqref{eq:mean} arises from the motion of the BCGs. If the BCGs are truly at the centre of their respective clusters, they are at rest and do not contribute to the Doppler effects. However, in practice, the BCG may not be located exactly at the bottom of their cluster's gravitational potential and hence have a residual velocity, encoded in the dispersion $\sigma^2_{\rm BCG}$ across the sample of stacked clusters. 

In addition to the shift of the mean in Eq.~\eqref{eq:mean}, we can also measure the width of the redshift difference distribution. In Appendix~\ref{app:variance}, we show that the variance, i.e.\ the square of the width, is given by
\begin{align}
\label{eq:var}
{\rm var}(\Delta z)_{R_\perp} =    \langle \left(\Delta z -\langle \Delta z \rangle_{R_\perp}\right)^2 \rangle_{R_\perp} = \sigma_{\rm los}^2 \left( R_\perp \right) + \sigma_{\rm BCG}^2  \equiv \sigma^2_{\rm tot}\, .
\end{align}
As expected, the width is governed by the linear Doppler effects of the cluster members and of the BCG, which can take any sign and thus enlarge the redshift distribution. The width is therefore larger than the shift by almost two orders of magnitude, i.e.~width $\sim \sigma_{\rm los}\sim 3\times 10^{-3}\gg$ shift $\sim \langle\Psi\rangle\sim \sigma_{\rm los}^2\sim 10^{-5}$.
In the following, we discuss how the width and the shift of the mean can be combined to test the validity of the weak equivalence principle.

\section{The test}

\begin{figure}
    \centering
    \includegraphics[width=0.62\linewidth]{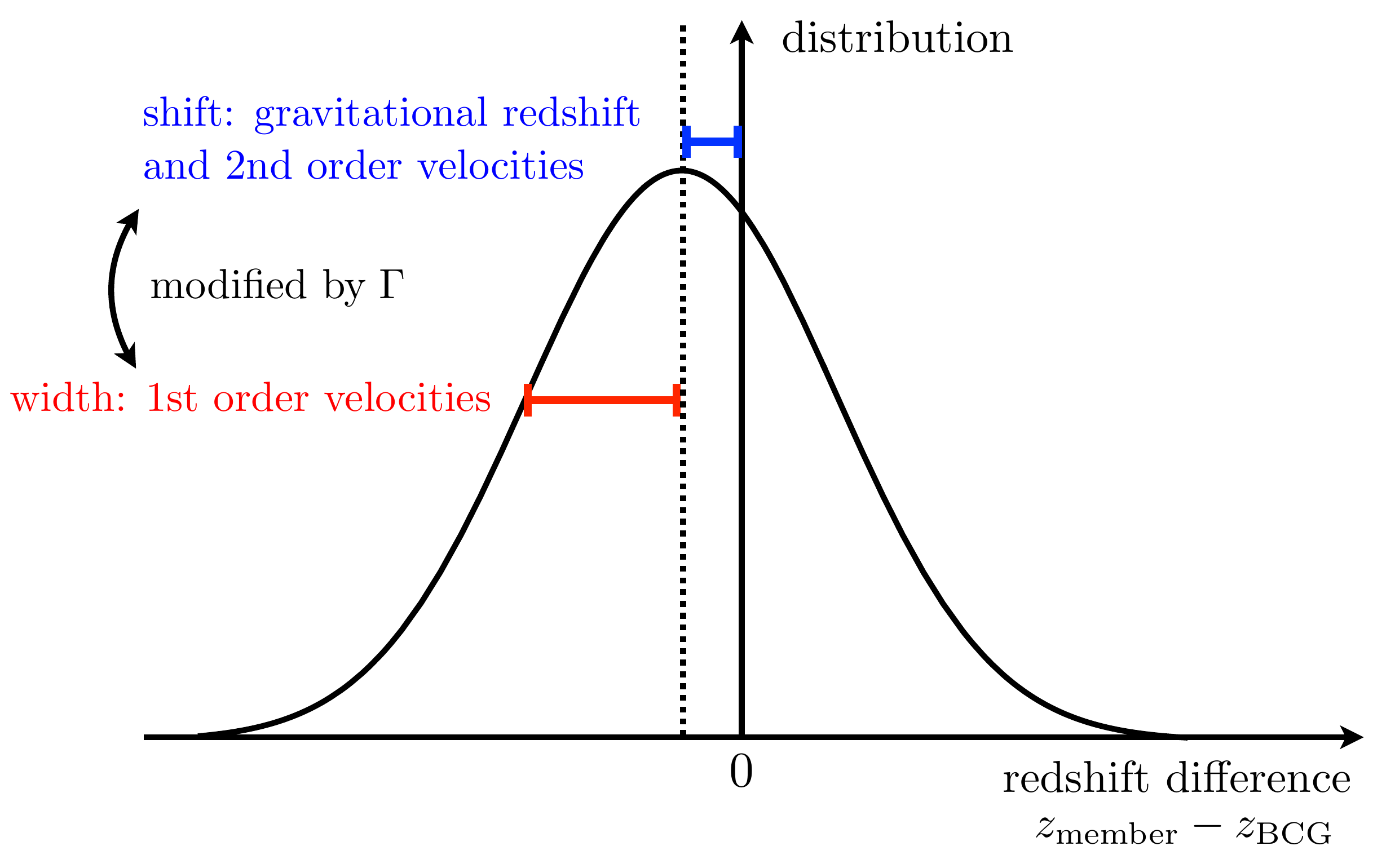}
    \caption{We measure the distribution of the redshift difference between the cluster members and the BCG. The width of the distribution is directly sensitive to the linear Doppler effect, which can take any sign, leading to the variance in Eq.~\eqref{eq:var}. Gravitational redshift and second-order Doppler effects, on the other hand, always have the same sign and shift the mean of the distribution away from zero following Eq.~\eqref{eq:mean}. A breaking of the weak equivalence principle, encoded in the fifth force $\Gamma$, would modify the relation between gravitational redshift and velocities and can thus be tested by combining the shift with the width.}
    \label{fig:test}
\end{figure}

The key idea of our test is to compare the shift of the redshift distribution $\langle \Delta z \rangle_{R_\perp}$ with its width ${\rm var}(\Delta z)_{R_\perp}$. These quantities depend on gravitational redshift and Doppler effects, which are linked through Jeans equation~\eqref{eq:Jeans}. Hence, if the weak equivalence principle is broken ($\Gamma\neq 0$), the relation between these two quantities will be altered and we can detect it, as depicted in Fig.~\ref{fig:test}.  

To perform this test, we express $\sigma^2_{\rm los}$ in the shift in Eq.~\eqref{eq:mean} and the variance in Eq.~\eqref{eq:var} as functions of the gravitational potential $\Delta\Psi$ using the modified Jeans equation~\eqref{eq:Jeans}:
\begin{align} 
 \label{eq:sigma2_los}
    \sigma^2_{\rm los} \left( R_\perp \right) &= \Sigma^{-1}\int dM \frac{dN_c}{dM} \int dr_e \ \rho_g^{\rm real}  \sigma^2_v   
    =
    - \Sigma^{-1}\int dM \frac{dN_c}{dM}  \int dr_e \ r_e \partial_{r_e} \left(  \rho_g^{\rm real} \sigma^2_v \right)  
  \\
    &= \Sigma^{-1}\left( 1+ \Gamma \right)  \int dM\frac{dN_c}{dM} \int dr_e \ r_e  \rho_g^{\rm real} \partial_{r_e} \Delta \Psi     \,.\nonumber
\end{align}
In the first line, we have applied integration by part, where the boundary terms vanish thanks to the isotropy of the stacked density profile,\footnote{The assumption of isotropy is not true in general for individual clusters, but stacking several with different masses tends to makes the resulting profile more isotropic.} implying $\rho(-r_e)\sigma_v^2(-r_e)=\rho(r_e)\sigma_v^2(r_e)$. In the second line, we have employed the Jeans equation to relate the velocity to the gravitational potential. This step relies on the fact that the derivative along the cylindrical coordinate $Z$ (the distance from the observer) coincides with the derivative along $r_e$, since these coordinates are aligned, see Fig.~2 in~\cite{DiDio:2025bff}.

Inserting~Eq.~\eqref{eq:sigma2_los} into Eqs.~\eqref{eq:mean} and~\eqref{eq:var}, we see that the shift and variance depend on the following quantities: the density profile of the clusters, which impacts $\rho_g^{\rm real}$ and $\Psi$ in the same way; the mass function of the cluster sample, $\frac{dN_c}{dM}$; the residual velocity dispersion of the BCGs, $\sigma_{\rm BCG}$; the function $\mathcal{R}$ that depends on the magnification bias and spectral index of the galaxies; and the function $\Gamma$, encoding the strength of a potential fifth force. To measure $\Gamma$, we need to express the other quantities in terms of a finite number of parameters, and determine whether there is enough information in the shift and the width of the redshift distribution to measure all of them. 

Following the approach of~\cite{Wojtak:2011ia}, we model the cluster mass function with a power law, $\frac{dN_c}{dM} \propto M^b$, and we adopt an NFW functional form for the density profile~\cite{Navarro:1995iw},
\begin{equation}
    \rho^{\rm real}_g  \propto \rho_{\rm NFW} = \frac{R_s}{R}\frac{\rho_0}{\left( 1 + \frac{ R}{R_s} \right)^2} \, .
\end{equation}
Here, $R$ is the radial distance from the cluster centre, related to $r_e$ and $R_\perp$ through $R^2=r_e^2+R_\perp^2$, while $R_s$ and $\rho_0$ are two free parameters that encode the properties of the halos. Moreover, we assume that general relativity is valid, so that this density profile is related to the gravitational potential $\Psi$ through the Poisson equation, yielding:
\begin{equation}\label{eq:Poisson_NFW}
    \Psi_{\rm NFW} \left( R \right) = \frac{4 \pi G R_s^3 \rho_0}{R} \log\left( \frac{R_s}{R+R_s}\right) \, .
\end{equation}
In practice, the profile can be expressed in terms of the concentration parameter $c_v \equiv \frac{R_v}{R_s}$, with $R_v$ the virial radius. Relating $R_v$ and $\rho_0$ to the mass $M$, as discussed in~\cite{DiDio:2025bff}, we can express $\Psi_{\rm NFW}$ and $\rho_{\rm NFW}$ in terms of $c_v$ only (since the signal is integrated over the mass $M$). With this setup, the shift and the width now only depend on the following five parameters: $\left\{ b,c_v, \mathcal{R} ,\Gamma, \sigma^2_{\rm BCG} \right\}$. We remark that while testing Jeans equation does not rely on a specific theory of gravity, as discussed in Section~\ref{sec:WEP}, here for simplicity we have assumed the validity of general relativity in relating the gravitational potential to the density profile in Eq.~\eqref{eq:Poisson_NFW}. This assumption could be relaxed, by allowing for modifications in the Poisson equation and constraining them together with our five parameters. However, this would likely require the inclusion of additional observations, such as gravitational lensing, in order to break degeneracies.

As we will show below, since the shift and the variance can be measured at different transverse separations $R_\perp$ from the cluster centre, there is enough information to constrain the five parameters simultaneously and measure $\Gamma$. In practice, however, some of the parameters can also be measured from independent observations, directly from the population of galaxies itself or from other signals, e.g.~gravitational lensing. Hence, we will consider different scenarios in our analysis, involving various priors on the parameters.

\section{Forecasting the detectability of a fifth force}

To determine the sensitivity of our test to a violation of the weak equivalence principle, we forecast the precision with which we can measure the parameter $\Gamma$ appearing in Eq.~\eqref{eq:Jeans}, encoding an additional fifth force acting solely on dark matter. We employ the Fisher formalism, which provides a powerful tool to predict the expected constraints on the parameters of interest given the specifications of the survey.

\subsection{Observables and Fisher formalism}
Our set of observables consists in measurements of the shift and the variance of the redshift distribution at various transverse separations $R_\perp$. For each bin in separation, labelled by $k$, we can construct estimators for the shift and the variance:
\begin{eqnarray}
        \mathcal{O}_k^{\rm est} =  \left( \begin{array}{c}
            \langle\Delta z \rangle_k^{\rm est}    \\
              {\rm var}(\Delta z)_k^{\rm est}  
    \end{array}\right) 
    {\renewcommand{\arraystretch}{1.5}
    \equiv \left( \begin{array}{c}
           \frac{1}{N_k} \sum\limits_{i=1}^{N_k} \Delta z_i (R^k_\perp)   \\
        \frac{1}{N_k-1} \sum\limits_{i=1}^{N_k} \big( \Delta z_i(R^k_\perp) -    \langle \Delta z \rangle_k^{\rm est} \big)^2
    \end{array}\right)}\, ,
\end{eqnarray}
where the sum runs over all cluster members in the separation bin centered on $R^k_\perp$, and $N_k$ is the total number of galaxies in that bin. These estimators are unbiased, i.e.~they converge to the theoretical prediction in Eqs.~\eqref{eq:mean} and \eqref{eq:var} in the limit of large $N_k$. Assuming that the individual redshift differences $\Delta z_i$ are identically and independently distributed,\footnote{This is strictly true for galaxies belonging to different clusters, which are independent from each other. Galaxies belonging to the same cluster have a small correlation through the gravitational field. In practice, however, thousands of clusters are stacked together in this type of measurement, typically with $\mathcal{O}(10)$ galaxies per cluster, such that this correlation is negligible.} we can compute the covariance of the estimator $\mathcal{O}_k^{\rm est}$,
\begin{eqnarray}
   \mathcal{C}_k = {\rm cov } \left( \mathcal{O}_k^{\rm est} \right) = \left( \begin{array}{cc}
        \frac{\sigma_{{\rm tot},k}^2}{N_k} \quad & \quad \frac{\gamma_{1,k} \sigma_{{\rm tot},k}^3}{N_k}  \\[5pt]
      \frac{\gamma_{1,k} \sigma_{{\rm tot},k}^3}{N_k}  \quad & \quad\sigma_{{\rm tot},k}^4 \left( \frac{2}{N_k-1} + \frac{\kappa_k}{N_k} \right) 
    \end{array} \right) \, .
\end{eqnarray}
As expected, the uncertainty on the shift is governed by the width of the distribution in the $k$-th bin, $\sigma_{{\rm tot},k}$. Since we treat the individual $\Delta z_i$ measurements as independent, this uncertainty scales as $1/\sqrt{N_k}$, i.e.~the more cluster members we observe, the better we can measure the shift and the width. If the redshift distribution were perfectly Gaussian, the uncertainty on the width would also be solely due to $\sigma_{{\rm tot},k}$, and the shift and the variance would be uncorrelated. Departures from Gaussianity are encoded in the skewness $\gamma_{1,k}$ and the kurtosis $\kappa_k$ of the distribution of $\Delta z_i$, which are defined as
\begin{eqnarray}
    \gamma_{1,k} &=& \frac{\mu_{3,k}}{\sigma^3_{{\rm tot},k}} = \frac{\mathbb{E} \left[ \left( \Delta z_i - \mathbb{E}\left[ \Delta z_i \right] \right)^3\right]}{\sigma^3_{{\rm tot},k}} \, ,
    \label{eq:skewness} \\
    \kappa_k &=& \frac{\mu_{4,k}}{\sigma^4_{{\rm tot},k}} -3 = \frac{\mathbb{E} \left[ \left( \Delta z_i - \mathbb{E}\left[ \Delta z_i \right] \right)^4\right]}{\sigma^4_{{\rm tot},k}} -3 \, . \label{eq:kurtosis}
\end{eqnarray}
where $\mathbb{E}$ indicates an expectation value and $\mu_{i,k}$ is the $i$-th central moment in the $k$-th bin. These quantities generate correlations between the shift and the width, as well as an increase of the uncertainty on the width. In the following, we will test the impact of these non-Gaussian features on the constraints, showing that they do not play an important role. 

We now have all the ingredients to forecast the uncertainty on the parameters. In the limit of large $N_k$, the central limit theorem ensures that the likelihood can be approximated by a Gaussian distribution. Moreover, we assume that the different $k$ bins are uncorrelated. This is not fully correct, since the gravitational potential and the velocity dispersion are indeed correlated at different separations. However, if the bins are sufficiently large, the correlations become small and can be neglected for the purpose of our forecast. In practice, when applying the measurement to data, the covariance between bins would be measured and included in the likelihood. 

For uncorrelated bins, the total Fisher matrix is simply the sum of the Fisher matrices in each bin~\cite{Tegmark:1997rp},
\begin{eqnarray} \label{eq:Fisher}
    F_{i j} = \sum_{k=1}^{N_{\rm bins}}\left\{\frac{\partial \mathcal{O}_k}{\partial \theta_i} \, \mathcal{C}^{-1}_{k} \frac{\partial\mathcal{O}_k}{\partial\theta_j}   + \frac{1}{2} {\rm tr} \left[ \mathcal{C}_k^{-1} \left( \frac{\partial\mathcal{C}}{\partial\theta_i}  \right)   \mathcal{C}_k^{-1} \left( \frac{\partial\mathcal{C}}{\partial\theta_j}  \right) \right]\right\} \, ,
\end{eqnarray}
where $\theta_i$ and $\theta_j$ run over the set of free parameters $\left\{ b,c_v, \mathcal{R} ,\Gamma, \sigma^2_{\rm BCG} \right\}$ and $k$ runs over the bins in $R_\perp$. In the following, we will consider four bins with upper limits: $[1.1,2.1,4.4,6]$ Mpc from the centre. We split the total number of galaxies $N$ in these four bins, assuming that their number density follows the surface density in Eq.~\eqref{eq:sigma_R}, leading to a relative number of galaxies per bin $N_k/N$ given by $\left\{0.40, 0.21, 0.27, 0.12 \right\}$. In Appendix~\ref{app:binning}, we show how the constraints vary if we change this distribution.

The expected 1$\sigma$ uncertainty on a given parameter $\theta_i$ is computed from the Fisher matrix as $\sigma_{\theta_i}=\sqrt{(F^{-1})_{ii}}$, where $F_{ij}$ is evaluated at the fiducial value of the parameters. In the following, we choose the fiducial combination $\left\{ b^{\rm fid}=-2.3,c_v^{\rm fid}=5.5, \mathcal{R}^{\rm fid}=6.4 ,\Gamma^{\rm fid}=0, (\sigma^{\rm fid}_{\rm BCG})^2= \left( 218 \ {\rm km/s} \right)^2 \right\}$, where we underline that we compute the uncertainty on the fifth force $\Gamma$ around 0. From the analytical expression of the Fisher matrix discussed in Section~\ref{sec:fifth_force_only}, we find that this uncertainty does not depend on the fiducial value when $\Gamma$ is the only free parameter.

\subsection{Constraints on the fifth force only}
\label{sec:fifth_force_only}
As a first step, we compute the uncertainty on $\Gamma$ by assuming that it is the only unknown parameter in the analysis. In this case, the uncertainty can be derived analytically, providing an intuitive understanding of the information encoded in the shift and the variance. In this derivation, we neglect the skewness and the kurtosis, which have a sub-dominant impact, as we will show in the next section. 

The derivative of the observable and of the covariance with respect to $\Gamma$ can be expressed as
\begin{eqnarray}
    \partial_\Gamma \mathcal{O}_k = \left( 
    \begin{array}{cc}
         \frac{3}{2} - \mathcal{R}  \\
          1
    \end{array}
    \right)  \sigma_{{\rm los},k}^2
    \hspace{1cm}\mbox{and}\hspace{1cm}
    \partial_\Gamma \mathcal{C}_k = 
    \left(
    \begin{array}{cc}
        \frac{\sigma^2_{{\rm los},k}}{N_k} & 0   \\
      0    &  4 \frac{\sigma_{{\rm los},k}^4}{N_k-1}
    \end{array}\right)\, ,
\end{eqnarray}
where all quantities are evaluated at their fiducial value. 
The Fisher element for $\Gamma$ then becomes
\begin{eqnarray} \label{eq:F_GammaGamma}
    F_{\Gamma \Gamma} 
    &=& \sum_{k=1}^{N_{\rm bin}}\left\{ \frac{1}{2} \left(N_k -1 \right) \frac{\sigma_{{\rm los},k}^4}{\sigma_{{\rm tot},k}^4} + N_k \left( \frac{3}{2} - \mathcal{R} \right)^2 \frac{\sigma^4_{{\rm los},k}}{\sigma^2_{{\rm tot},k}}  + \frac{1}{2} \left( 4 \frac{\sigma^8_{{\rm los},k}}{\sigma^8_{{\rm tot},k}} + \frac{\sigma^4_{{\rm los},k}}{\sigma^4_{{\rm tot},k}}\right)\right\}
    \nonumber \\
    &\overset{\left(\sigma^2_{\rm BCG} \rightarrow 0\right)}{=}&  \frac{1}{2} \left(N -1 \right) + \sum_{k=1}^{N_{\rm bin}}N_k \left( \frac{3}{2} - \mathcal{R} \right)^2 \sigma^2_{{\rm los},k} + \frac{5}{2}\, .
\end{eqnarray}
To achieve an analytical understanding of the different terms, we show the limit when the velocity dispersion of the BCGs goes to zero in the second line. In practice, $\sigma^2_{\rm los}/\sigma^2_{\rm tot}$ ranges between $0.8$ and $0.9$, depending on $R_\perp$. 

We see that three terms contribute to the Fisher information on $\Gamma$ in Eq.~\eqref{eq:F_GammaGamma}: the first two terms respectively encode the information coming from the variance of the distribution, ${\rm var}(\Delta z)_{R_\perp}$, and the shift, $\langle \Delta z \rangle_{R_\perp}$, while the last term is due to the second contribution to the Fisher matrix in Eq.~\eqref{eq:Fisher}. When the observed number of galaxies $N$ is large (current measurement typically involve $10^5$ galaxies), this last term is strongly sub-dominant and can thus be neglected. By contrast, the information derived from both the variance and the shift scales linearly with the number of sources $N$, leading to a total uncertainty on $\Gamma$ scaling as $\sigma_\Gamma\sim 1/\sqrt{N}$. 

\begin{figure}
    \centering
                \includegraphics[width=0.49\linewidth]{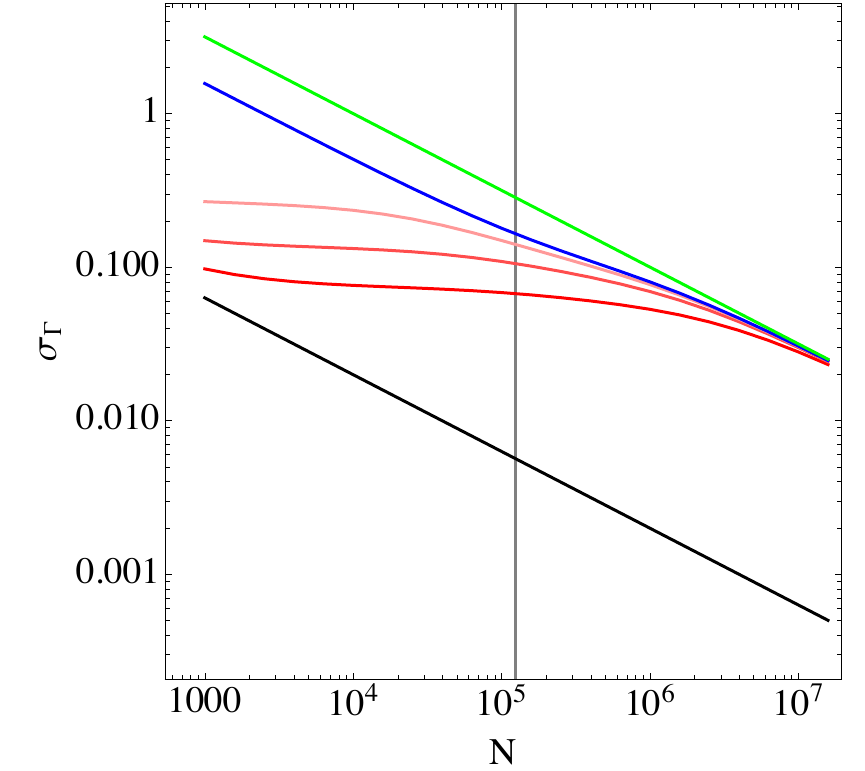}                  
    \caption{1$\sigma$ uncertainty on $\Gamma$ as a function of the number of galaxies $N$, with different assumptions about the other parameters. The black line shows the result where all parameters are fixed except $\Gamma$, while the green line corresponds to the case without any prior on any parameter. Intermediate lines all have a 20\% Gaussian prior on $\mathcal{R}$ and $\sigma_{\rm BCG}^2$ and no prior on $c_v$, but they differ in terms of $b$, which is strongly degenerated with $\Gamma$: the blue line has no prior on $b$, while the light to dark red lines respectively have priors of (20\%, 10\%, 5\%) on $b$. The vertical line corresponds to the number of galaxies considered in \cite{Wojtak:2011ia}.
    }
    \label{fig:Fisher_Gamma_only}
\end{figure}

Comparing the information from the variance in the first term with that from the shift in the second term, we see that the latter is strongly sub-dominant, since $\sigma^2_{\rm los} \sim \left(500 \, {\rm km/s}\right)^2 \simeq 3 \times 10^{-6}$ and the prefactor $\left(\tfrac{3}{2} - \mathcal{R}\right) \sim \mathcal{O}(1$–$10)$. Hence, if we only measured the shift of the mean from zero in an ideal situation where all parameters are perfectly known, we would detect a violation of the weak equivalence principle of the order of 35\% with around $125'000$ galaxies.\footnote{As we see from Eq.~\eqref{eq:Jeans}, $\Gamma$ modifies the Jeans equation by adding a new force in addition to the gravitational interaction. As such, a value of $\Gamma$ of 0.35 leads to a breaking of the pure gravitational interaction, i.e.~a violation of the weak equivalence principle, of 35\%.} This is roughly the number of galaxies observed in~\cite{Wojtak:2011ia}, which we consider as our baseline case.

By adding the measurement of the width of the distribution, i.e.~of the variance, we strongly improve the constraints, reaching a precision of 0.6\%  on $\Gamma$ with $125'000$ galaxies, as can be seen from the black line in Fig.~\ref{fig:Fisher_Gamma_only}. This strong constraint on $\Gamma$ arising from the variance reflects the fact that the depth of the gravitational potentials is perfectly known when all parameters other than $\Gamma$ are fixed. Consequently, by measuring the width, which is directly proportional to the galaxy velocity dispersion $\sigma_{\rm los}$, we can tightly constrain the presence of a fifth force that would modify the link between the depth of the potential and the velocity at which galaxies fall inside it. In this idealised scenario, each galaxy provides a measurement with a signal-to-noise ratio of order unity coming from the width. In practice, it is of course unrealistic to assume that the masses and profiles of the clusters are perfectly known. Hence, in the following, we derive the constraints on $\Gamma$ letting the other parameters vary. We will see that in these realistic scenarios, even though the shift has a much smaller signal-to-noise ratio than the width, it plays a crucial role in constraining $\Gamma$ since it breaks degeneracies in the parameter space.

\subsection{Constraints on all parameters and degeneracies}

Intuitively, we understand that the width is characterised by a perfect degeneracy between the parameter $\Gamma$ and the masses and profiles of the clusters. Galaxy velocities can indeed be modified either by changing the masses and profiles of the clusters, or by introducing a non-zero $\Gamma$. Measurements of the shift are expected to break these degeneracies by probing the depth of the gravitational potential. To quantify these degeneracies, the capability of the shift to break them, and the resulting constraints on $\Gamma$, we numerically compute the Fisher matrix and the uncertainties by varying all five parameters $\left\{ b,c_v, \mathcal{R} ,\Gamma, \sigma^2_{\rm BCG} \right\}$. 

We neglect the second term in the Fisher matrix in Eq.~\eqref{eq:Fisher}, since it is strongly sub-dominant in the large $N$ limit, as we showed for $\Gamma$ in Eq.~\eqref{eq:F_GammaGamma}.\footnote{This contribution to the Fisher matrix arises from the Gaussian likelihood assumption, which treats the shift and variance as independent. This strictly holds only in the limit $N \to \infty$. As emphasised in~\cite{Carron:2012pw}, it is generally safer to fix the covariance at its fiducial value and drop this term, thereby avoiding the inclusion of spurious information arising from the dependence of the covariance on the parameters of the underlying model.} As before, we consider four bins in $R_\perp$ in the forecasts. While considering more numerous, thinner bins may help break degeneracies thanks to the sensitivity  of some parameters to $R_\perp$, this would require accounting for the covariance between neighboring bins, which are no longer independent when they are close to each other. Since we do not know this covariance, we restrict our forecasts to four bins, in order not to artificially increase the constraining power. When working with data, it is instead necessary to estimate the covariance between the bins and optimise the size and number of bins. 

We present our results in Fig.~\ref{fig:Fisher_Gamma_only}. The green line shows the constraints on $\Gamma$ as a function of the number of galaxies, keeping all other parameters free and marginalising over them. The constraints are degraded by roughly two orders of magnitude with respect to the case where $\Gamma$ is the only free parameter discussed in the previous section (black line). For example, we only reach a precision of 28\% with $N = 125'000$ galaxies. This degradation is mainly due to the degeneracy between $\Gamma$, $b$ and $c_v$.

\begin{figure}
    \centering
            \includegraphics[width=0.49\linewidth]{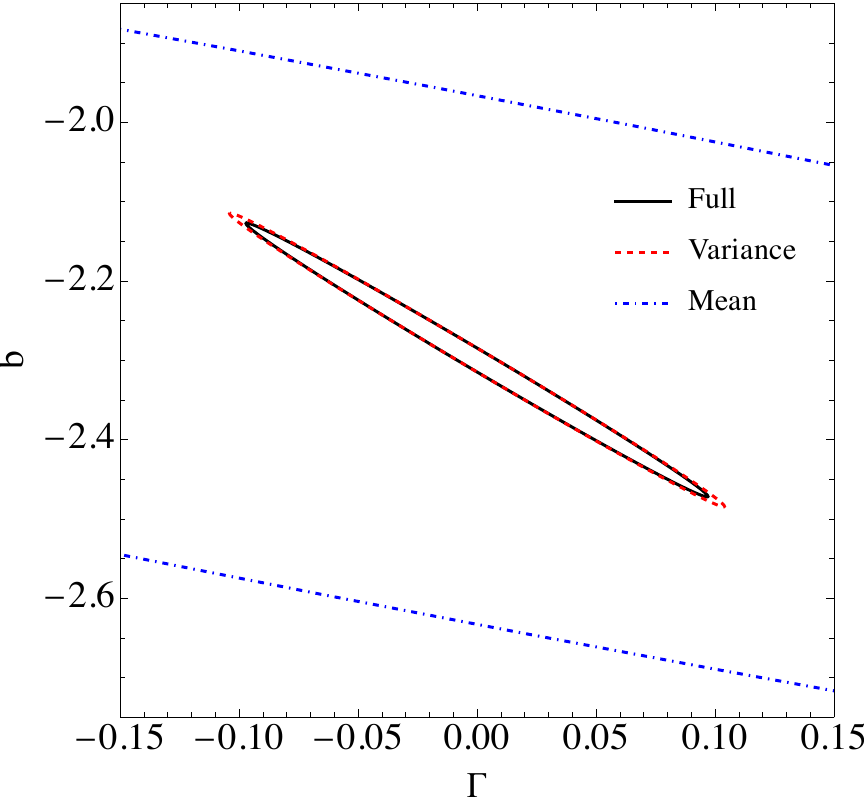}
        \includegraphics[width=0.49\linewidth]{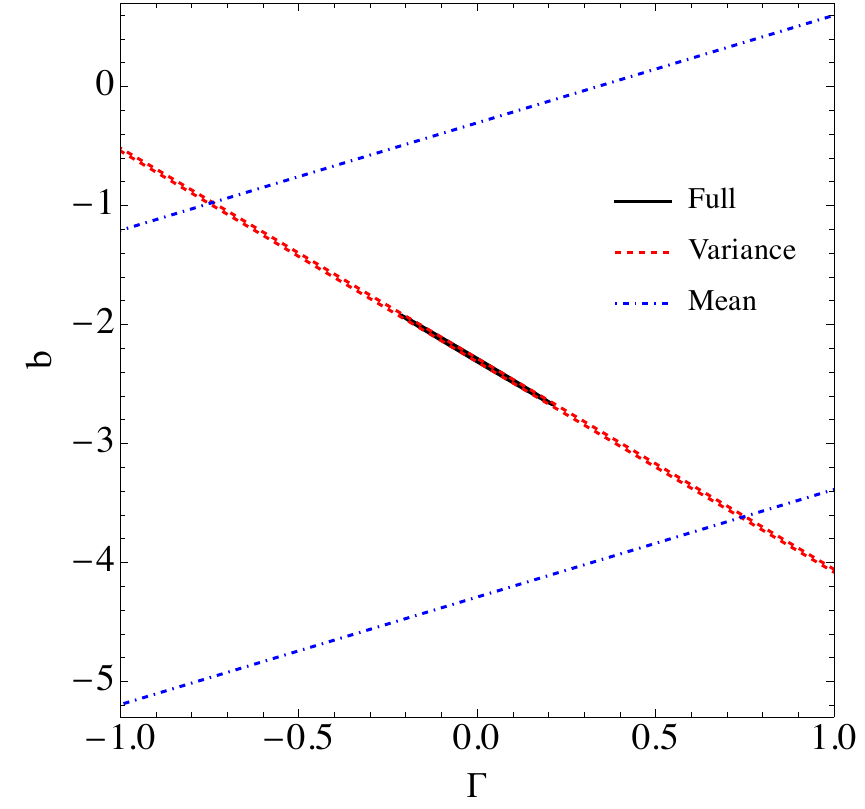} 
    \caption{Joint 1$\sigma$ constraints on $b$ and $\Gamma$ obtained from the variance only (red), the shift only (blue) and the combination of the two (black). In the left panel, we keep all the other parameters fixed at their fiducial values, while in the right panel, we also vary $c_v$ and marginalise over it.
    }
    \label{fig:Fisher_Gamma_b}
\end{figure}

This degeneracy is illustrated in Fig.~\ref{fig:Fisher_Gamma_b}. In the left panel, we show the joint constraints on $b$ and $\Gamma$, keeping all other parameters fixed, for $N=125'000$ galaxies. We compare the constraints from the variance alone (red), the shift alone (blue) and the combination of the two (black). Using the variance only, which is sensitive to galaxy velocities, we see that $b$ and $\Gamma$ are quite degenerate. This is expected, since varying $b$ modifies the mass distribution of the clusters, which in turn changes the galaxy velocities. For example, decreasing $b$ decreases the number of clusters with high mass (since the mass function becomes steeper for more negative $b$), which in turn decreases the galaxy velocities. This change can be reabsorbed into a positive $\Gamma$, which increases the galaxy velocities.

The degeneracy is however partially broken by the measurement of the variance at several separations $R_\perp$, which are affected by a change in $b$ and in $\Gamma$ in a different way. As a result, the constraints on $\Gamma$ from the variance alone are still relatively tight, of the order of 8\% (versus 0.6\% when $b$ is fixed). In this scenario, adding gravitational redshift measured through the shift has almost no impact on the constraints. Even though gravitational redshift is not affected by the same degeneracy between $b$ and $\Gamma$ as the velocities, its signal-to-noise ratio is much smaller than that of the variance, so that it does not improve the constraints.  

The situation completely changes when also leaving $c_v$ free, which impacts the profile of the gravitational potential. The results in this case are shown in the right panel of Fig.~\ref{fig:Fisher_Gamma_b}. In this scenario, there is enough freedom in $c_v$ and $b$, to {\it completely mimic} the impact of a fifth force on the galaxy velocities at all observed separations $R_\perp$, as can be seen from the red contour. This is not surprising: if neither the masses of the clusters nor their profile are known, we cannot constrain $\Gamma$ by only looking at the way galaxies move in the clusters. Therefore, measuring gravitational redshift through the shift becomes crucial, as this has the capacity to break the degeneracies. Indeed, when $b$ and $c_v$ vary, the width and the shift are modified in a consistent way. For example, the gravitational potential can become deeper, leading to both a stronger gravitational redshift signal and larger velocities. Such an effect is intrinsically different from that of a fifth force arising from a nonzero $\Gamma$, which breaks the link between the gravitational potential and the velocities. For a given gravitational potential -- corresponding to a fixed gravitational redshift signal -- one can for example have larger velocities due to a positive $\Gamma$, making dark matter move faster. As a consequence, measuring both the shift and the width is necessary to disentangle the effect of $\Gamma$ from the combined effect of $b$ and $c_v$, as we can see from the black contour in Fig.~\ref{fig:Fisher_Gamma_b}.

In practice, the parameters $b, c_v, \mathcal{R}$ and $\sigma_{\rm BCG}$ are not completely unknown and can be measured with independent methods. For example, the magnification bias $s_b$ and the spectral index $\alpha$ that appear in $\mathcal{R}$ can be directly measured from the population of galaxies, see e.g.~\cite{Wang:2020ibf,Maartens:2021dqy}. The cluster mass function and hence $b$ can be constrained from weak gravitational lensing, from X-ray emissions or from the thermal Sunyaev–Zeldovich effect, see e.g.~\cite{Vikhlinin:2005mp,Vikhlinin:2008cd,vonderLinden:2014haa,Planck:2015koh,Wen:2015uqa}. Moreover, the velocity dispersion of the BCGs can be estimated by comparing their redshift with the mean redshift of the clusters, which can be obtained by averaging over all members of the clusters, see e.g.~\cite{2014ApJ...797...82L}. To incorporate these additional measurements in our forecasts, we include priors on the different parameters, in order to assess their impact on the uncertainty on $\Gamma$. 

\begin{figure}
    \centering
    \includegraphics[width=0.49\linewidth]{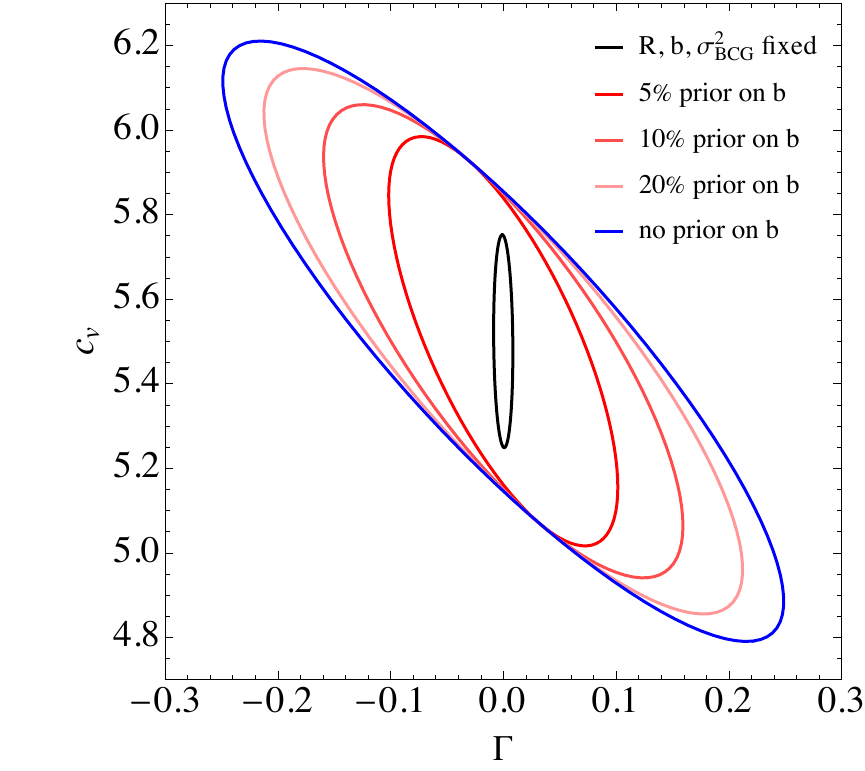}
    \caption{Joint 1$\sigma$ constraints on $c_v$ and $\Gamma$ for $N=125'000$. When the other parameters are kept fixed (black line), $c_v$ and $\Gamma$ are not degenerate. The other lines include a 20\% Gaussian prior on $\mathcal{R}$ and $\sigma_{\rm BCG}^2$. We show the case with no prior on $b$ (blue line) and the cases with priors on $b$ of \{5\%, 10\%, 20\%\} in shades from dark to light red. This illustrates how the constraints on $\Gamma$ degrade when $c_v$ and $b$ are both unknown.}
    \label{fig:Fisher_Gamma}
\end{figure}

In Fig.~\ref{fig:Fisher_Gamma_only}, we plot the results for various prior configurations. The blue line corresponds to a 20\% prior on $\mathcal{R}$ and $\sigma_{\rm BCG}$ and no prior on $b$ and $c_v$. With respect to the case without priors, this tightens the constraints by almost a factor 2 when considering $125'000$ galaxies. The three red lines include the same prior on $\mathcal{R}$ and $\sigma_{\rm BCG}$, as well as an additional prior on $b$ of (20\%, 10\%, 5\%), respectively. Such priors lead to a precision of 7-14\% on $\Gamma$ for our baseline number of galaxies. When increasing this number, the constraints become tighter and the priors on $b$ become less and less relevant. Note that we do not impose a prior on $c_v$ in these results. This parameter can in principle be measured from lensing or X-ray emission close to the centre of the clusters, allowing for a reconstruction of the density profile of the clusters. However, such measurements are challenging, hence we choose a conservative approach where $c_v$ is solely constrained from our measurements, i.e.\ from the dynamics of the cluster, and thus no prior is imposed on it. In Fig.~\ref{fig:Fisher_Gamma}, we show the degeneracy between $c_v$ and $\Gamma$ for different priors on $b$. When $b$ is perfectly known, $c_v$ and $\Gamma$ are not degenerate, while relaxing the priors on $b$ gradually increases the degeneracies. In Appendix \ref{app:full_corner_plot}, for completeness, we show the joint constraints on all pairs of parameters. 

In our setup, a large fraction of galaxies is situated close to the center, since we have assumed that the galaxy number density follows that of the matter density described by the NFW profile, leading to $N_k/N\in\left\{0.40, 0.21, 0.27, 0.12 \right\}$ for the four bins. As a consequence, most of the constraining power on $\Gamma$ arises from the first two bins. For example, with a prior of $10 \%$ on $b$ and of $20\%$ on $\sigma^2_{\mathrm{BCG}}$ and $\mathcal{R}$ (corresponding to the central red line in Fig.~\ref{fig:Fisher_Gamma_only}), the innermost two bins provide a $12\%$ constraint on $\Gamma$ with 125'000 galaxies. This is only tightened to $11\%$ by including the third bin, while the fourth bin does not bring any additional information. 

In practice, however, the observed distribution of galaxies is determined by the specifications of the survey, in particular concerning the precision of the redshift measurement for the cluster members, and may not follow the density profile. To assess the impact of the distribution on the constraints, in Appendix~\ref{app:binning}, we calculate the optimal distribution of galaxies that gives the tightest constraints on $\Gamma$. We find that the optimal configuration involves a large fraction of galaxies in the most distant bin and a small fraction in the first bin. This is due to two effects: on the one hand, the signal-to-noise ratio of gravitational redshift increases with the distance to the center, leading to stronger constraints on $\Gamma$ from the most distant bin; on the other hand, the various parameters have a different impact at different $R_\perp$, so that having measurements at very different $R_\perp$ helps breaking degeneracies. The importance of the first bin with respect to the fourth one directly depends on the choice of priors, and we find that the tighter the priors, the fewer galaxies are needed in the first bin. This optimal distribution improves the constraints by $30-60\%$ with respect to our baseline configuration, depending on the choice of priors. 

We also checked the impact of higher-order moments encoding the deviation of the redshift distribution from Gaussianity. To do so, we extracted the skewness and kurtosis in Eqs.~\eqref{eq:skewness}--\eqref{eq:kurtosis} from the redshift distributions plotted in~\cite{Wojtak:2011ia}, finding $\gamma_1=\{-0.016, -0.017, 0.003, 0.003\}$ and $\kappa=\{0.530, 0.995, 1.235, 0.881\}$. We then repeated the Fisher analysis including them in the covariance. We found that the constraints on $b$ and $\Gamma$ are degraded by 0.3\% only, while those on $c_v$ by 5\%. This minor degradation can be traced to the additional covariance between the shift and the variance generated by the skewness. 

Lastly, we studied the importance of including second-order Doppler effects in the theoretical model for the shift. As discussed previously, second-order Doppler effects contribute to the shift with a similar amplitude to that of gravitational redshift. Hence, neglecting them would not be consistent and would lead to a biased measurement of $\Gamma$. In Appendix~\ref{app:bias}, we compute the amplitude of this bias for $\Gamma$ and for $b$. We find that if only $\Gamma$ and $b$ are left free and all the other parameters are fixed to their fiducial values, $\Gamma$ is biased by roughly 1$\sigma$. Such a large bias is not acceptable, demonstrating the importance of properly accounting for second-order Doppler effects. When all parameters are left free, the bias on $\Gamma$ is slightly smaller. This is expected, since in this case more parameters can be tuned to accommodate for the wrong modelling of the signal. In addition, the uncertainty on $\Gamma$ significantly increases, so that the bias becomes negligible in this case.

\section{Comparison with other tests}

Since its first measurement in clusters, gravitational redshift has been used to test models beyond $\Lambda$CDM, in particular departures from general relativity~\cite{Wojtak:2011ia,eBOSS:2021ofn,Rosselli:2022qoz}. The idea behind such tests is that the depth of gravitational potentials is affected by gravity modifications, leading to a stronger or weaker gravitational redshift signal than in general relativity. Thus, constraints on gravity modifications can be obtained by comparing the measured gravitational redshift signal with predictions in various theories of gravity. However, the challenge of such tests is that they require independent knowledge of the masses of the clusters. The mass is used to infer the gravitational potential $\Psi$ in a given theory of gravity, which is then compared with the value of $\Psi$ that is directly measured from gravitational redshift, in order to look for deviations. 

In~\cite{Wojtak:2011ia}, the authors determined the cluster masses from the galaxy velocities measured from the width of the redshift distribution. In this procedure, they employed the Poisson equation to relate the cluster density profile to its gravitational potential. However, in the presence of gravity modifications, the Poisson equation is modified, so that it cannot be used to make this link. As a consequence, it is not possible to test modified gravity by combining measurements of gravitational redshift and of the mass inferred from the galaxy velocities, as already pointed out in~\cite{Zhao:2012gxk}. On a deeper level, this limitation is due to the fact that modifying gravity changes both the galaxy velocities and the gravitational potential $\Psi$, but it does not change the link between the two, as along as the weak equivalence principle remains valid, i.e.~the modifications affect all components involved (baryons and dark matter) in the same way. This is explicit in our derivation of the modified Jeans equation in Appendix~\ref{app:distribution}, which does not rely on any theory of gravity. Therefore, it is not possible to test theories beyond general relativity by comparing these two quantities without additional external information~\cite{Zhao:2012gxk,Kaiser:2013ipa}. By contrast, the combination of galaxy velocities and the gravitational potential provides a natural test of the weak equivalence principle, as demonstrated in this work. 

Performing consistent tests of modified gravity is still feasible, but only provided that we have independent knowledge of the cluster masses. In~\cite{eBOSS:2021ofn}, the mass distribution of the clusters was inferred from the X-ray properties of the clusters. The density profile of the clusters was then obtained from a scaling relation between the masses and the concentration parameter appearing in the NFW profile~\cite{Dutton:2014xda}. The resulting profile was used to predict the gravitational redshift signal in $f(R)$ theories, which were then compared with the measurements. Such a procedure is valid as long as three conditions are met: 1) the relation between X-ray measurements and the inferred masses is not altered by gravity modifications; 2) the density profile is well-described by an NFW functional form in the modified gravity scenario; and 3) the scaling relation between masses and concentration parameters is not modified. These three conditions may not hold for all theories beyond general relativity, as shown for example in~\cite{Terukina:2013eqa,Wilcox:2015kna}.

In~\cite{Rosselli:2022qoz}, the authors estimated the cluster masses and concentration parameter explicitly for $f(R)$ theories, finding results that are on average about 15\% higher than in general relativity. This shows the importance of consistently accounting for modifications of gravity in all steps of the analysis, not only in the prediction of $\Psi$. While the differences found in~\cite{Rosselli:2022qoz} are small compared with uncertainties in the measurements of the gravitational redshift signal, they may be relevant for future surveys.

Our test is intrinsically different from these previous ones, since it is not designed to probe departures from general relativity, intended as fundamental modifications to the Einstein equations. The aim of our approach is rather to test for a breaking of the weak equivalence principle, which modifies the relation between velocities and gravitational potential. As such, we do not need to directly infer the masses of the clusters, but we rather parametrise the mass distribution and the density profile with a set of free parameters that we constrain together with the fifth force. However, it is important to stress that a fifth force could in principle modify the density profile, not only through a different value of $c_v$ (that would be captured by our test), but by introducing deviations from the NFW functional form. If a fifth force were to be detected, the validity of this assumption should be checked, for example by running simulations that include a fifth force. In this way, it would be possible to determine the true functional form of the density profile, and in case it differs from the NFW one, the analysis could be repeated with a new parametrised form. Consequently, our test can provide a robust detection of a non-zero $\Gamma$-- in the sense that if there is no fifth force, the NFW profile is correct and it cannot artificially lead to $\Gamma \neq 0$ -- but the specific non-zero value of $\Gamma$ that we detect may be impacted by an incorrect profile. 

\section{Conclusions}
\label{sec:conclusions}

In this work, we have proposed a test of the weak equivalence principle on cluster scales. By parametrising deviations from the Vlasov equation, we have shown that the dynamics of galaxies in clusters, described by the Jeans equation, are directly sensitive to a fifth force parameter $\Gamma$. Using stacked measurements of the redshift difference between cluster members and the BCG, we have exploited the shift and the width of the distribution, demonstrating how these observables can be combined to constrain possible departures from the weak equivalence principle for dark matter.

We have performed a Fisher analysis forecasting the constraints as a function of the number of galaxies in the sample of stacked clusters. Our results indicate that most of the constraining power comes from the width of the redshift difference, which encodes the velocity distribution of the cluster members. The shift, although less informative on its own, plays a crucial role in breaking degeneracies, in particular between the cluster mass function and density profile and the parameter $\Gamma$, thus distinguishing between the depth of the gravitational potential and the effect of a fifth force on dark matter. We have shown that the constraints on $\Gamma$ range between $7\%$ and $14\%$ for a sample of $10^5$ galaxies, depending on whether the prior on the mass function power-law parameter $b$ is set to $5\%$ or $20\%$. This demonstrates the feasibility of our test with current measurements of gravitational redshift, and in a forthcoming paper we will apply it to the data of~\cite{Wojtak:2011ia,Sadeh:2014rya,eBOSS:2021ofn,Rosselli:2022qoz}.
The newest  generation of galaxy surveys, like DESI and Euclid, will detect a significantly higher number of clusters and of galaxies, decreasing the uncertainty on $\Gamma$ to a few percent, as shown in Fig.~\ref{fig:Fisher_Gamma}.

Our test ideally complements the one performed in~\cite{Grimm:2025pmq}, which searched for a fifth force on cosmological scales by combining galaxy velocities with gravitational lensing measured from cosmic shear. The measurements constrained a positive fifth force to be smaller than 7\% of the gravitational interaction and a negative one smaller than 21\%. Our test has the strong advantage that it does not rely on the validity of general relativity, since it directly probes the link between the velocity dispersion and the distortion of time. In contrast, the test in~\cite{Grimm:2025pmq} only holds provided that general relativity is the correct theory of gravity. This limitation is due to the fact that gravitational redshift, and thus the distortion of time, has not yet been measured at cosmological scales, so that it could not be employed for the test in \cite{Grimm:2025pmq}. Moreover, our test probes a highly complementary regime to the cosmological one by accessing cluster scales, where dark matter is densely distributed and governs both the dynamics and the depth of the gravitational potential. In particular, this provides sensitivity to shorter-range dark matter interactions, which could have a negligible impact on cosmological scales, but a strong impact close to the centre of clusters. Hence, performing this test with current and upcoming data will provide crucial information on the properties of dark matter and answer the fundamental question of the validity of the weak equivalence principle in galaxy clusters.

\section*{Acknowledgements}
We are grateful to Fabian Schmidt for precious comments that improved the quality of this manuscript. We thank Ruth Durrer, Øyvind Christiansen, David Mota, Charlie Mpetha, Veronika Oehl, Simon White and Hans Winther for insightful discussions. We acknowledge funding from the European Research Council (ERC) under the European Union’s Horizon 2020 research and innovation program (Grant No. 863929; project title ``Testing the law of gravity with novel large-scale structure observables''). SC is supported by a Postdoc.Mobility Fellowship of the Swiss National Science Foundation (SNSF), project number P500PT\_230281.

\begin{appendices}

\section{Modified Jeans equation} \label{app:distribution}

We derive the Jeans equation in the presence of dark matter interactions given in Eq.~\eqref{eq:Jeans} in the main text. We assume a perturbed Friedmann Universe, with the metric in the Newtonian gauge given by
\begin{equation}\label{eq:perturbed_FLRW_metric}
    g_{\mu \nu} dx^\mu dx^\nu = a^2(\eta) \left[ -\left( 1 + 2 \Psi \right) d\eta^2 + \left( 1 - 2 \Phi \right) d\mathbf{x}^2\right] \, ,
\end{equation}
where $a$ is the scale factor, $\eta$ denotes conformal time, and $\Phi$ and $\Psi$ are the two gravitational potentials. On the scale of clusters, the gravitational potentials remain small, of the order of $10^{-5}$, while velocities are typically two orders of magnitude larger, of the order of $10^{-3}$, and the density perturbations are of order unity or larger. To account for these different orders of magnitude, we derive the Jeans equation in the weak field expansion. More precisely, from the Einstein and conservation equations, we obtain the following parametric relations between density, velocities and gravitational potentials, 
\begin{equation}
    \delta \sim \epsilon^{-1}_\HH v \sim \epsilon_\HH^{-2} \Psi \sim \epsilon_\HH^{-2} \Phi\, ,
\end{equation}
where $\epsilon_\HH = \HH/k$ is the weak field expansion parameter, which is much smaller than 1 inside the horizon, i.e.~when $k\gg\HH$. In the following, we treat $\delta$ non-perturbatively, and we consistently expand $v,\Psi$ and $\Phi$ at the desired order in $\epH$.

To obtain the Jeans equation, we start by deriving the Vlasov equation, which describes the evolution of the distribution function of galaxies $f(x^\mu,p^\mu)$. Here, $x^\mu$ represents the coordinates and $p^\mu=mu^\mu$ the 4-momentum of galaxies, with $m$ the mass of galaxies and $u^\mu$ their four-velocities. The function $f$ represents the number density of galaxies in phase space with respect to the invariant measure $d\mu=2\sqrt{|g|}d^4x\sqrt{|g|}d^4p \delta_D(p^2+m^2)$, where $g$ is the determinant of the metric in Eq.~\eqref{eq:perturbed_FLRW_metric}, the Dirac delta function $\delta_D$ ensures that $p^2\equiv p^\mu p_\mu=-m^2$, and the factor 2 is a useful convention. Integrating over $p^0$ to evaluate $\delta_D$, we obtain 
\begin{align}
d\mu_m= \frac{\sqrt{|g|}}{|p_0|} d^3p \sqrt{|g|}d^4x\, .   
\end{align}
The energy-momentum tensor is then calculated by integrating over momenta,
\begin{align}
\label{eq:T}
T^{\mu\nu}=\int d^3p\frac{\sqrt{|g|}}{|p_0|}p^\mu p^\nu f(x^\alpha,p^i)\,.     
\end{align}
If galaxies move on geodesics, i.e.~in the case where there are no additional interactions acting on dark matter, $f$ obeys the Vlasov equation through Liouville's theorem,
\begin{align}
\label{eq:Df}
 \frac{Df}{D\tau}\equiv\frac{\partial f}{\partial x^\alpha}\frac{dx^\alpha}{d\tau}+\frac{\partial f}{\partial p^i}\frac{dp^i}{d\tau}=0\, ,   
\end{align}
where $\tau$ denotes the proper time of galaxies. Using the geodesic equation, this leads to
\begin{align}
\label{eq:ffixedp}
p^\alpha\frac{\partial f}{\partial x^\alpha} -\frac{\partial f}{\partial p^i}\Gamma^i_{\mu\nu}p^\mu p^\nu=0\, .
\end{align}
From this equation, we can derive the Euler and continuity equations at lowest order in $\epH$. At this order, $p^\mu=\frac{m}{a}(1,\tilde{v}^i)$, where $\tilde{v}^i$ is the galaxy peculiar velocity, which we denote with a tilde here, to differentiate it from the mean velocity below. Following~\cite{Durrer:2008eom}, it is convenient to express Eq.~\eqref{eq:Df} in terms of the momentum $\hp^i\equiv a p^i=m\tilde{v}^i$. To do so, we need to rewrite the derivatives at fixed value of $p^i$ in terms of derivatives at fixed value of $\hp^i$:
\begin{align}
\frac{\partial f}{\partial x^0}\Big|_{ p}&=\frac{\partial f}{\partial x^0}\Big|_{ \hp}+\frac{\partial f}{\partial\hp^j}\frac{\partial \hp^j}{\partial x^0}\Big|_{p}=\frac{\partial f}{\partial x^0}\Big|_{ \hp}+\HH\hp^j\frac{\partial f}{\partial\hp^j}\, ,\label{eq:f1}\\
\frac{\partial f}{\partial x^i}\Big|_{ p}&=\frac{\partial f}{\partial x^i}\Big|_{ \hp}+\frac{\partial f}{\partial\hp^j}\frac{\partial \hp^j}{\partial x^i}\Big|_{p}=\frac{\partial f}{\partial x^i}\Big|_{ \hp}\, ,\label{eq:f2}\\
\frac{\partial f}{\partial p^i}&=a\frac{\partial f}{\partial \hp^i}\, .\label{eq:f3}
\end{align}
In the following, we only use derivatives at fixed $\hp$, so that we drop the symbol $|_{\hp}$ for simplicity. Inserting Eqs.~\eqref{eq:f1}--\eqref{eq:f3} into~\eqref{eq:ffixedp}, we obtain
\begin{align}
\label{eq:eqf_final}
\frac{\partial f}{\partial\eta}-\HH\hp^i\frac{\partial f}{\partial \hp^i}+\tilde{v}^i\frac{\partial f}{\partial x^i}-m\partial^i\Psi\frac{\partial f}{\partial\hp^i}=0\, .    
\end{align}

The zeroth moment of this equation is the continuity equation, which encodes the conservation of mass. This can obtained by integrating Eq.~\eqref{eq:eqf_final} over momentum and multiplying it by $m$, yielding
\begin{align}
\partial_\eta\rho+\partial_i(\rho v^i)+3\HH\rho=0\, .    
\end{align}
Here, we have introduced the energy density $\rho$ and the mean velocity $v^i$, which are obtained by integrating the distribution function over momentum. From Eq.~\eqref{eq:T}, we obtain at lowest order in $\epH$ 
\begin{align}
\rho&=a^2 T^{00}=m\int d^3\hp f \, ,\\
\rho v^i&=a^2T^{i0}=m\int d^3\hp\, \tilde{v}^i f\, .
\end{align}

The first moment of the Vlasov equation yields the Euler equation, which can be obtained by multiplying Eq.~\eqref{eq:eqf_final} by $m \tilde{v}^i$ and integrating over momentum, yielding
\begin{align}
\partial_\eta(\rho v^i)+4\HH\rho v^i +\rho\partial^i\Psi+\partial_j\left[\rho(\sigma^{2\,ij}+v^i v^j) \right]=0\, .   
\end{align}
Here, the velocity dispersion $\sigma^{2\,ij}$ is defined through
\begin{align}
m\int d^3\hp\, \tilde{v}^i\tilde{v}^j f=\rho(\sigma^{2\,ij}+v^i v^j)\, .    
\end{align}

Any additional non-gravitational interactions acting on dark matter would modify Eq.~\eqref{eq:ffixedp}. Here, we aim to express the modifications with a phenomenological approach, in order to test for their presence without specifying a model for the additional interactions. To do so, we use the fact that additional interactions affecting dark matter would modify the geodesic equation,
\begin{equation}
\frac{d p^i}{d\tau}+\Gamma^i_{\mu\nu}p^\mu p^\nu = \mbox{corrections}\, .
\label{eq:geo}
\end{equation}
The form of the corrections depends on the type of interaction and can be derived explicitly within specific models. For example, in the case of a scalar field $\phi$ mediating a fifth force through a conformal coupling, the right-hand side of Eq.~\eqref{eq:geo} acquires a term proportional to the gradient of the scalar field $\partial^i \phi$ (see Eq.~(2.9) in~\cite{Bonvin:2018ckp}). In the presence of a force mediated by a vector field $A_\mu$, various modifications appear, proportional to $p^i, A^i$ and $(\partial_\mu A^i-\partial^i A_\mu)p^\mu$ (see Eq.~(2.21) in~\cite{Bonvin:2018ckp}). Lastly, for a force mediated by a tensor field $\hat{g}_{\mu\nu}$, the modifications are of the form $\hat{\Gamma}^i_{\mu\nu}p^\mu p^\nu$, with $\hat{\Gamma}$ the Christoffel symbols of the second metric. To account for all these types of interactions, mediated by a scalar, vector or tensor field, we add all possible terms up to second-order in $p^\mu p^\nu$ to Eq.~\eqref{eq:geo}. Terms involving more $p^\mu$'s may appear in the case of interactions mediated by higher-order spin fields, but we do not consider such scenarios here. In full generality, the perturbations in the scalar, vector and tensor fields can be related to the metric potentials $\Phi$ and $\Psi$ and their derivatives via the Einstein equations. We can therefore consistently construct all possible modifications at order $\epsilon_\HH$, taking into account that $\Psi,\Phi$ are of order $\epH^2$, that $p^i$ is of order $\epH$, and that a gradient brings an enhancement of $\epH^{-1}$. With this, we obtain
\begin{align}
&p^0\frac{\partial f}{\partial x^0}+p^i\frac{\partial f}{\partial x^i}-\frac{\partial f}{\partial p^i}\Gamma^i_{00}(p^0)^2-C_\Psi\frac{\partial f}{\partial p^i}\partial^i\Psi(p^0)^2-C_\Phi\frac{\partial f}{\partial p^i}\partial^i\Phi(p^0)^2 -2\frac{\partial f}{\partial p^i}\Gamma^i_{0j}p^0p^j\nonumber\\
&-C_p\frac{\partial f}{\partial p^i} \delta^i_j p^0p^j-C_{\partial\Psi}\frac{\partial f}{\partial p^i}\partial^i\partial_j\Psi p^0p^j
-C_{\partial\Phi}\frac{\partial f}{\partial p^i}\partial^i\partial_j\Phi p^0p^j- C_{\partial^2\Psi}\frac{\partial f}{\partial p^i}\partial^i\partial_j\partial_k\Psi p^j p^k\nonumber\\
&- C_{\partial^2\Phi}\frac{\partial f}{\partial p^i}\partial^i\partial_j\partial_k\Phi p^j p^k=0\, . \label{eq:fmod}
\end{align}
The quantities $C_\alpha$ are phenomenological parameters that encode the strength of the different types of interactions. In the following, we assume that they do not depend on the momentum $p$, which is the case for interactions mediated by a scalar, vector or tensor field, as discussed above (see also explicit expressions in~\cite{Bonvin:2018ckp}). However, this excludes dark matter models with a momentum-dependent interaction, such as self-interacting dark matter, where the collision of dark matter particles depends on momentum, see e.g.~\cite{Tulin:2017ara}. Note that the $p^0$'s in Eq.~\eqref{eq:fmod}, which are of order $\epH^0$, are explicitly written in order to obtain an equation that is homogeneous in the mass $m$. They could alternatively be incorporated in the $C_\alpha$ coefficients. Taking the zeroth moment of Eq.~\eqref{eq:fmod}, we obtain the modified continuity equation,
\begin{align}
\partial_\eta\rho+\partial_i(\rho v^i)+3(\HH+C_p)\rho+\rho(C_{\partial\Psi}\Delta\Psi+C_{\partial\Phi}\Delta\Phi)
+2C_{\partial^2\Psi}\rho v^i\partial_i\Delta\Psi+2C_{\partial^2\Phi}\rho v^i\partial_i\Delta\Phi=0\, ,\label{eq:continuity_mod}
\end{align}
while the first moment gives the modified Euler equation,
\begin{align}
&\partial_\eta(\rho v^i)+4(\HH+C_p)\rho v^i +(1+C_\Psi)\rho\partial^i\Psi+C_\Phi\rho\partial^i\Phi+\partial_j\left[\rho(\sigma^{2\,ij}+v^i v^j) \right]\nonumber\\
&+C_{\partial\Psi}\rho \big(v^i\Delta\Psi+v^j\partial_j\partial^i\Psi\big)+C_{\partial\Phi}\rho \big(v^i\Delta\Phi+v^j\partial_j\partial^i\Phi\big)\nonumber\\
&+C_{\partial^2\Psi}\rho\Big[(\sigma^{2\,jk}+v^jv^k)\partial_j\partial_k\partial^i\Psi+2(\sigma^{2\,ij}+v^iv^j)\partial_j\Delta\Psi \Big]\nonumber\\
&+C_{\partial^2\Phi}\rho\Big[(\sigma^{2\,jk}+v^jv^k)\partial_j\partial_k\partial^i\Phi+2(\sigma^{2\,ij}+v^iv^j)\partial_j\Delta\Phi \Big]=0 \, .\label{eq:euler_mod}
\end{align}
As in the standard scenario, the continuity equation~\eqref{eq:continuity_mod} links the evolution of the density $\rho$ to the velocity $v^i$, and in the Euler equation~\eqref{eq:euler_mod}, the evolution of $v^i$ depends on the velocity dispersion $\sigma^{ij}$. To solve this system, it is not sufficient to only consider these two equations, since they depend on the evolution of the velocity dispersion, which is unknown. Instead, it would be necessary to derive the hierarchy of moments of the Vlasov equation, where each moment is sourced from the following one and all equations depend on the modified coefficients $C_\alpha$. One could solve the system for specific forms of these coefficients, truncating the hierarchy at some order.

This is however not the goal of our test. In our case, we aim to test for the presence of non-zero values of the $C_\alpha$ coefficients in a model-independent way -- i.e.~without specifying their form -- by comparing measurements of $\Psi$ and of the velocity dispersion. To obtain a relation that can be used for this purpose, we first insert the modified continuity equation~\eqref{eq:continuity_mod} into the modified Euler equation~\eqref{eq:euler_mod}, yielding
\begin{align}
&\rho\partial_\eta v^i+(\HH+C_p)\rho v^i +(1+C_\Psi)\rho\partial^i\Psi+C_\Phi\rho\partial^i\Phi+\big[\partial_j(\rho\sigma^{2\,ij})+\rho v^j\partial_j v^i\big]\nonumber\\
&+C_{\partial\Psi}\rho v^j\partial_j\partial^i\Psi +C_{\partial^2\Psi}\rho\Big[(\sigma^{2\,jk}+v^jv^k)\partial_j\partial_k\partial^i\Psi+2\sigma^{2\,ij}\partial_j\Delta\Psi \Big]\nonumber\\
&+C_{\partial\Phi}\rho v^j\partial_j\partial^i\Phi+C_{\partial^2\Phi}\rho\Big[(\sigma^{2\,jk}+v^jv^k)\partial_j\partial_k\partial^i\Phi+2\sigma^{2\,ij}\partial_j\Delta\Phi \Big]=0 \, .\label{eq:euler_mod_final}
\end{align}
We then switch to cylindrical coordinates $\{R_\perp, \varphi, Z \}$ and adopt the simplifying assumption that the clusters of galaxies used in the measurement are virialised, so that the mean velocity vanishes in the clusters' rest frame and we can set the mean velocity $v^i$ to zero in Eq.~\eqref{eq:euler_mod_final}. Moreover, we assume isotropic clusters, so that the velocity dispersion is the same in all directions, i.e.~$\sigma^{2\,ij}=\sigma_v^2 e_x^i e_x^j+\sigma_v^2 e_y^i e_y^j+\sigma_v^2 e_z^i e_z^j$, where $\{\mathbf{e}_x,\mathbf{e}_y,\mathbf{e}_z\}$ are orthonormal vectors. In terms of the cylindrical unit vectors, we obtain $\sigma^{2\,ij}=\sigma_v^2 e_\perp^i e_\perp^j+\sigma_v^2 e_\varphi^i e_\varphi^j+\sigma_v^2 e_Z^i e_Z^j$.\footnote{Here, $\mathbf{e}_z=\mathbf{e}_Z$, and we use $Z$ in the following to avoid confusion with the redshift $z$.} Lastly, under isotropy, the gravitational potentials $\Phi$ and $\Psi$ are independent of the angle $\varphi$. With this, Eq.~\eqref{eq:euler_mod_final} becomes
\begin{align}
\partial_Z(\rho\sigma_v^2)+\left(1+C_\Psi+\frac{C_\Phi}{1+\varpi}\right)\rho\partial_Z\Psi  
+3\left(C_{\partial^2\Psi}+\frac{C_{\partial^2\Phi}}{1+\varpi}\right)\rho\sigma_v^2\left(\partial_Z^2+\partial^2_\perp+\frac{1}{R_\perp}\partial_\perp \right)\partial_Z\Psi=0\, ,\label{eq:Jeans_mod}
\end{align}
where we have related $\Phi$ to $\Psi$ with the gravitational slip $\varpi$ via $\Psi = \left( 1 +\varpi\right) \Phi$. Note that $\varpi=0$ in general relativity. 

We see that by measuring $\Psi$ from the shift of the redshift distribution and $\rho\sigma_v^2$ from the width, we can test for the presence of non-zero $C_\Psi, C_\Phi, C_{\partial^2\Psi}$ and $C_{\partial^2\Phi}$. In the following, we concentrate on classes of models with no energy exchange, i.e.~where the continuity equation~\eqref{eq:continuity_mod} is not modified. This implies in particular that $C_{\partial^2\Psi}=C_{\partial^2\Phi}=0$, and Eq.~\eqref{eq:Jeans_mod} takes the form of the usual Jeans equation with a modified term,
\begin{align}
\label{eq:Jeans_final}
\partial_Z(\rho\sigma_v^2)+(1+\Gamma)\partial_Z\Psi=0\, ,   
\end{align}
where
\begin{align}
\Gamma= C_\Psi+\frac{C_\Phi}{1+\varpi}\, .
\end{align}
The parameter $\Gamma$ encodes the strength of an additional fifth force acting on dark matter, which impacts the velocity dispersion of the galaxies inside the cluster. Eq.~\eqref{eq:Jeans_final} is the fundamental relation that underlies the test presented in the main text. There, the quantities $\Psi$ and $\sigma_v^2$ are measured from the shift and the width of the redshift distribution, respectively, while $\rho$ is assumed to take an NFW functional form. Hence, this equation can be used to obtain constraints on $\Gamma$.

We remark that the derivation above does not include the impact of a fraction of baryons respecting the equivalence principle. We leave a detailed computation for future work, but we expect this to lead to a rescaling of the parameter $\Gamma$ in Eq.~\eqref{eq:Jeans_final} as a function of the fraction of dark matter in the clusters, similarly to the impact on linear scales \cite{Castello:2022uuu}. This would slightly change the measured value of $\Gamma$, but, most importantly, it cannot lead to a spurious detection of a non-zero value for $\Gamma$. The presence of baryons indeed cannot mimic the effect of a fifth force in the Jeans equation, since they obey the weak equivalence principle. Hence, our test is robust to the impact of baryonic effects.

\section{Variance of the redshift difference} \label{app:variance}

We compute the variance of the redshift difference at the lowest order in the weak field expansion, which will turn out to be our order of interest, i.e.~$\epsilon_{\mathcal{H}}^2$. We will see that, at this order, the variance is purely sensitive to the velocity dispersion along the line of sight. Since the mean of the redshift difference in Eq.~\eqref{eq:mean} is non-zero only from second order in the weak field expansion, it does not contribute to the variance at the lowest order. Thus, we have 
\begin{equation} \label{eq:B6}
   {\rm var}(\Delta z)_{R_\perp}= \langle \left(\Delta z  -   \langle \Delta z  \rangle_{R_\perp}\right)^2 \rangle_{R_\perp} = \frac{\int dM\frac{dN_c}{dM}  \int d^3 \left( \Delta v \right) f \left(\boldsymbol{\Delta} \bv \right) \int dz \ n_g \left( \bn, z, F\ge F_* \right) \left(\Delta z \right)^2}{\int dM \frac{dN_c}{dM} \int dz \ n_g \left( \bn, z, F\ge F_* \right) } \, ,
\end{equation} 
where (see \cite{DiDio:2025bff} for more details)
\begin{equation}
    \Delta z = \HH_e r_e - \Delta v_\parallel + \mathcal{O} \left( \epsilon_\HH^2 \right) \, .
\end{equation}
Here, $\Delta v_\parallel$ encodes the impact of the linear Doppler effect on the redshift difference, and $\HH_e r_e$ is due to the background cosmological evolution between the time of emission of the BCG and that of the galaxy members. The contributions at order $\epsilon_\HH^2$, arising from gravitational redshift and second-order Doppler effects, generate higher-order contributions to the variance, and are therefore neglected here.
Since $\Delta z$ is linear in $\epsilon_\HH$, its contribution to Eq.~\eqref{eq:B6} is already at second order, meaning that we can keep all other terms at zeroth-order. This leads to
\begin{eqnarray} \label{eq:B8}
    {\rm var}(\Delta z)_{R_\perp} &=&\frac{\int dM\frac{dN_c}{dM} \int d^3 \left( \Delta v \right) f \left(\boldsymbol{\Delta} \bv \right) \int dr_e \ \rho_g^{\rm real} \left( \eta_e, r_e, F\ge F_* \right) \left(\Delta z \right)^2}{\int dM \frac{dN_c}{dM} \int dr_e \ \rho_g^{\rm real} \left( \eta_e, r_e, F\ge F_* \right)  } 
    \nonumber \\
    &=&\frac{\int dM \frac{dN_c}{dM}  \int dr_e \ \rho_g^{\rm real} \left( \eta_e, r_e, F\ge F_* \right) \left(\HH_e^2 r_e^2 + \sigma^2_v \right)}{\int dM \frac{dN_c}{dM} \int dr_e \ \rho_g^{\rm real} \left( \eta_e, r_e, F\ge F_* \right)  } +\mathcal{O} \left( \epsilon_\HH^3 \right) \, ,
\end{eqnarray}
where we have used
 \begin{equation}
     \int d^3 \Delta v \ f\left(\boldsymbol{\Delta} \bv  \right) \left(\Delta v_\parallel\right)^2 = 
          \int d^3 \Delta v \ f\left(\Delta v  \right) \left(\Delta v_\parallel\right)^2 = \sigma^2_v \, ,
 \end{equation}
and the mixed term $\HH_e r_e \Delta v_\parallel$ vanishes once integrated over the velocity dispersion due to parity. We remark that the variance also contains a contribution from the evolution of the background. As discussed in \cite{DiDio:2025bff}, this contribution can however be removed by fitting for a linear slope in the redshift distribution. Thus, we obtain
\begin{equation}
      {\rm var}(\Delta z)_{R_\perp} = \sigma^2_{\rm los} \equiv  \frac{\int dM\frac{dN_c}{dM}  \int dr_e \ \rho_g^{\rm real} \left( \eta_e, r_e, F\ge F_* \right) \sigma^2_v}{\int dM\frac{dN_c}{dM} \int dr_e \ \rho_g^{\rm real} \left( \eta_e, r_e, F\ge F_* \right)  }  +\mathcal{O} \left( \epsilon_\HH^3 \right) \, .
\end{equation}
As expected, the lowest-order contribution to the variance is purely given by the linear Doppler effect, which here appears squared following from the definition of the variance. In this derivation, we have neglected the velocity of the BCG, assuming that it is at rest at the bottom of the potential. As discussed in detail in~\cite{DiDio:2025bff}, this is not the case in practice, and it is necessary to add a term accounting for the velocity dispersion of the BCGs, i.e.
\begin{equation}
{\rm var}(\Delta z)_{R_\perp} = \sigma^2_{\rm los}+\sigma^2_{\rm BCG}+\mathcal{O} \left( \epsilon_\HH^3 \right)\, .    
\end{equation}

\section{Optimal bin-weighting}
\label{app:binning}

In our analysis, we have assumed that the density of galaxies in clusters follow that of the surface density, for which we have adopted an NFW profile. In practice, however, the galaxies with a sufficiently precise redshift determination to measure gravitational redshift may not follow such a profile in a specific survey. Here, we thus investigate how the constraints on $\Gamma$ vary with the galaxy distribution. More precisely, we determine the optimal distribution of galaxies minimising the constraints on $\Gamma$, and we calculate the improvement in the constraints with respect to our baseline results. We consider four bins as in our baseline analysis. Assuming the bins to be independent, we can write the Fisher matrix as
\begin{eqnarray}
    \boldsymbol{F} \left(f_1 , f_2 , f_3, f_4 \right)  = \sum_{i=1}^4 f_i \boldsymbol{F}_i \,,
\end{eqnarray}
where $\boldsymbol{F}_i$ denotes the Fisher matrix for the $i$-th bin and $f_i$ denotes the fraction of galaxies in that bin. We then minimise $\sqrt{(\boldsymbol{F}^{-1})_{\Gamma \Gamma}}$ with respect to $f_i$ for a fixed total number of galaxies $N$, under the conditions that
\begin{eqnarray}
    \sum_{i=1}^4 f_i = 1  \quad {\rm and} \quad 0\le f_i \le 1 \,.
\end{eqnarray}

\begin{figure}
    \centering
        \includegraphics[width=0.32\linewidth]{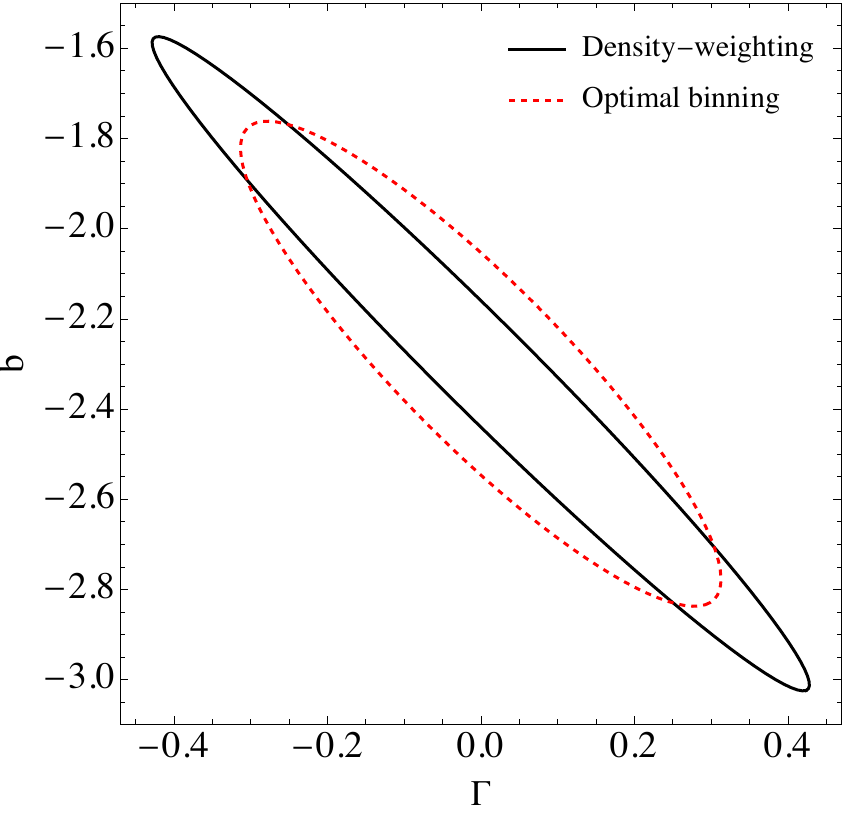}
    \includegraphics[width=0.32\linewidth]{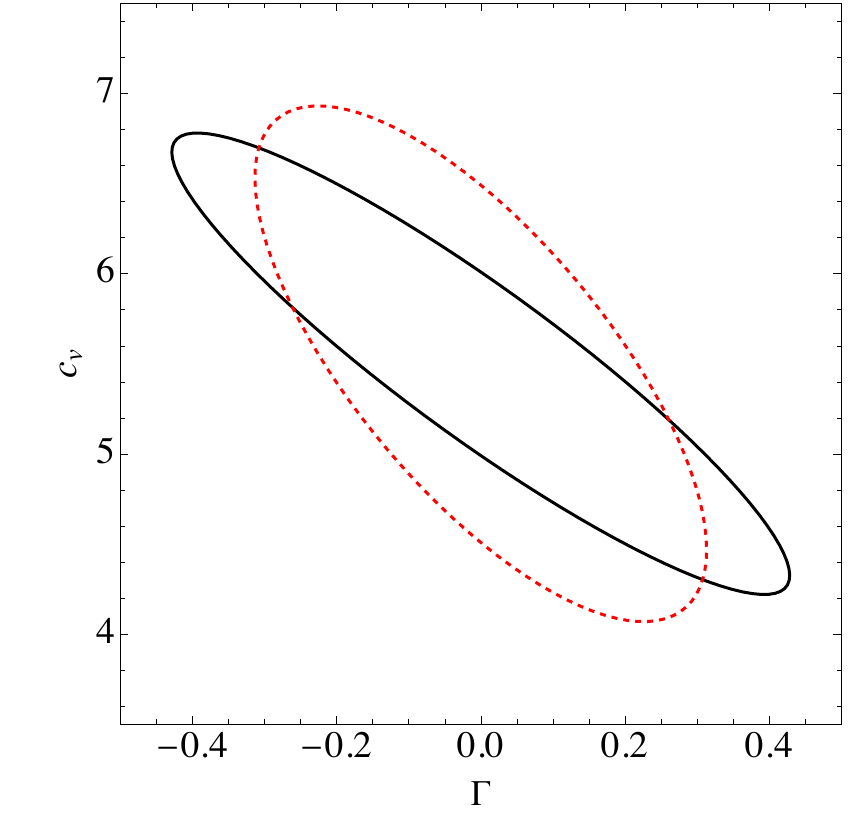}
     \includegraphics[width=0.32\linewidth]{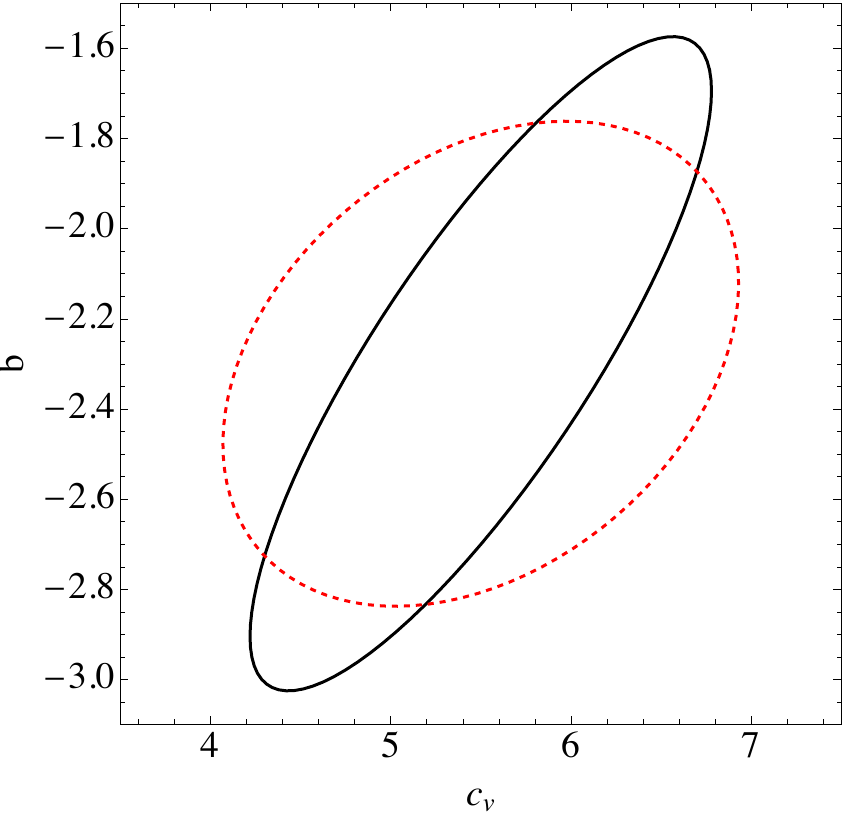}
    \caption{Comparison of the joint 1$\sigma$ constraints with the density-weighted binning (solid black line) and the optimal binning (red dashed line). All other parameters are marginalised over.}
    \label{fig:Fisher_Binning}
\end{figure}

In the case where there is no prior on the other parameters, we find that the optimal bin weighting is given by
\begin{eqnarray}
    \left\{ f_1, f_2, f_3, f_4\right\} = \left\{0.37, 0.05, 0., 0.59\right\} \, .
\end{eqnarray}
For our reference number of galaxies $N=125'000$, such a binning would improve the constraint on $\Gamma$ by a factor $1.64$, going from $0.36$ with the density-weighted binning used in the main text to $0.22$. We see that it is important to have galaxies well separated in $R_\perp$. In terms of signal-to-noise ratio, the gravitational redshift difference is larger for galaxies that are far away from the centre, since they are located much higher in the gravitational potential than the BCG. On the other hand, the uncertainty on the shift, which is due to $\sigma_{\rm los}$, decreases with separation. Hence, the gravitational redshift signal is better measured at large separation, while the signal-to-noise ratio of the width is independent of separation, since both the signal and the uncertainty have the same scaling in $R_\perp$, proportional to $\sigma^2_{\rm los}$. Given this, one could naively expect that $\Gamma$ is better constrained when all galaxies are situated in the outer bins. This is however not the case, due to the degeneracies between parameters. The sensitivity of $\Psi$ and $\sigma_{\rm los}$ to the various parameters scales with $R_\perp$ in a different way, and having widely separated $R_\perp$ bins turns out to crucially help to break degeneracies. This is illustrated in Fig.~\ref{fig:Fisher_Binning}, where we see that the degeneracy between $\Gamma$ and $b$ is significantly reduced in the optimal binning with respect to our baseline case. 

We also consider the case with $20\%$ priors on $\mathcal{R}$ and $\sigma_{\rm BCG}^2$, for which we find that the optimal binning is given by
\begin{eqnarray}
    \left\{ f_1, f_2, f_3, f_4\right\} = \left\{0.08, 0, 0, 0.92\right\} \, .
\end{eqnarray}
In this case, the priors already provide significant degeneracy-breaking power, such that the best constraints are obtained with only two bins, with most galaxies located in the outskirts of the cluster where the gravitational redshift signal is larger.
The optimal binning reduces the marginalised error on $\Gamma$ from $0.14$ with the density-weighted binning to $0.11$, i.e.\ an improvement of a factor $1.27$.

We underline that this binning does not optimise the total Fisher information. As shown in Fig.~\ref{fig:Fisher_Binning}, the marginalised error on $\Gamma$ is reduced by the optimal binning primarily by minimising the uncertainty along the degeneracy direction with $b$ and also by alleviating the degeneracy with $c_v$. However, this binning choice leads to weaker constraints on $c_v$, even though its degeneracy with $b$ is also mitigated.

\section{Bias due to incorrect modelling of the shift}
\label{app:bias}

\begin{figure}
    \centering
    \includegraphics[width=0.5\linewidth]{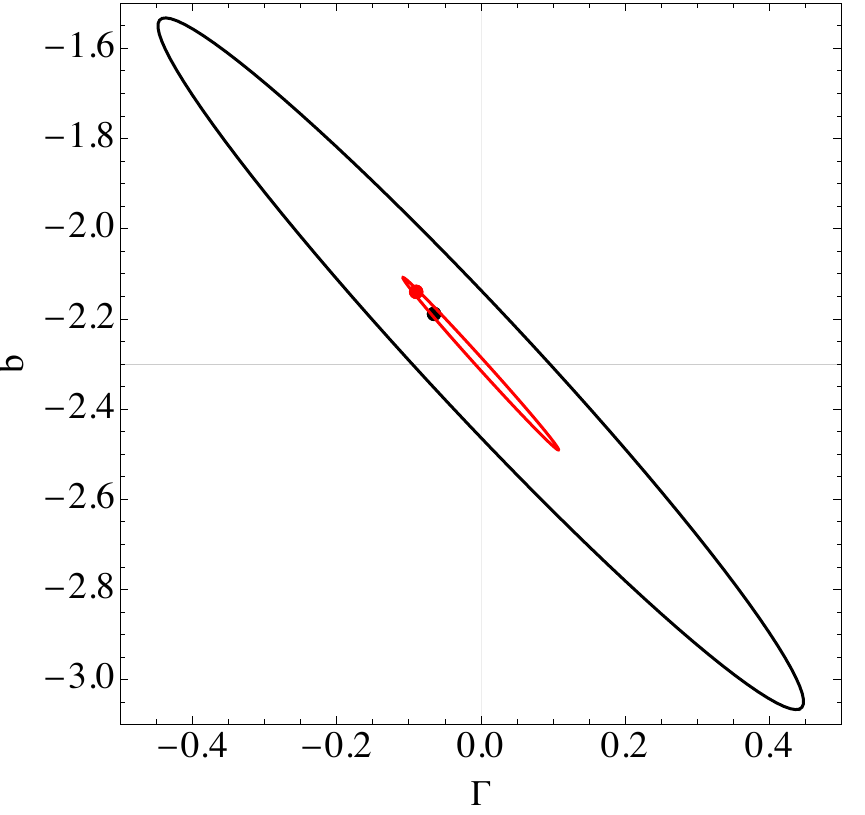}
    \caption{We show the best fit points and joint 1$\sigma$ contour for the parameters $b$ and $\Gamma$, obtained when neglecting the second-order Doppler effects in the shift. The black point and contour are related to the case where all other parameters are kept fixed, while the red ones refer to the case where we marginalise over the other parameters. The fiducial value is at the intersection of the grey lines.}
    \label{fig:bias}
\end{figure}

In this appendix, we study the bias introduced by an incorrect modeling of the shift $\langle \Delta z \rangle_{R_\perp}$ in the parameter inference. In particular, we are interested in understanding how the constraints on the fifth force parameter $\Gamma$ are biased under the assumption that the shift of the redshift distribution is due to the gravitational redshift effect only. We therefore extend Eq.~\eqref{eq:mean} to
\begin{align}
\label{eq:mean_epsilon}
\langle \Delta z \rangle_{R_\perp} = -  \langle \Delta \Psi \rangle_{R_\perp}  + \epsilon \left[ \left( \frac{3}{2} - \mathcal{R}\right) \sigma_{\rm los}^2(R_\perp)  - \frac{7}{2} \sigma_{\rm BCG}^2 \right]\, ,    
\end{align}
where the correct expression is recovered with $\epsilon=1$.
Following~\cite{Taylor:2006aw,Knox:1998fp,Heavens:2007ka,Kitching:2008eq}, we can estimate the bias in the parameters through
\begin{align}
    \Delta \theta_\alpha = \sum_\beta \left( F^{\theta \theta} \right)^{-1}_{\alpha \beta} F_\beta^{\theta \epsilon} \
\end{align}
where $\theta$ denotes the parameters of the model  and
\begin{align}
    F^{\beta \epsilon}_\beta = \sum_{k \ell} \frac{\partial \mathcal{O}_k}{\partial \theta_\beta} \mathcal{C}^{-1}_{k \ell} B_\ell\, .
\end{align}
Here,
\begin{align}
    \mathbf{B} = \left(\begin{array}{c}  \left( \frac{3}{2} - \mathcal{R}\right) \sigma_{\rm los}^2(R_\perp)  - \frac{7}{2} \sigma_{\rm BCG}^2  \\0 \end{array} \right) 
 \end{align}
with all the parameters evaluated at their fiducial values. The results for the parameters $b$ and $\Gamma$ are plotted in Fig.~\ref{fig:bias}. When all other parameters are kept fixed, these parameters are biased by almost 1$\sigma$, showing the importance of properly including the second-order Doppler contaminations in the modelling of the shift. When the other parameters are left free, the bias remains almost the same, but since the uncertainty increases significantly, its impact becomes negligible.

\section{Constraints on the full parameter space}\label{app:full_corner_plot}

\begin{figure}[h!]
    \centering
    \includegraphics[width=0.9\linewidth]{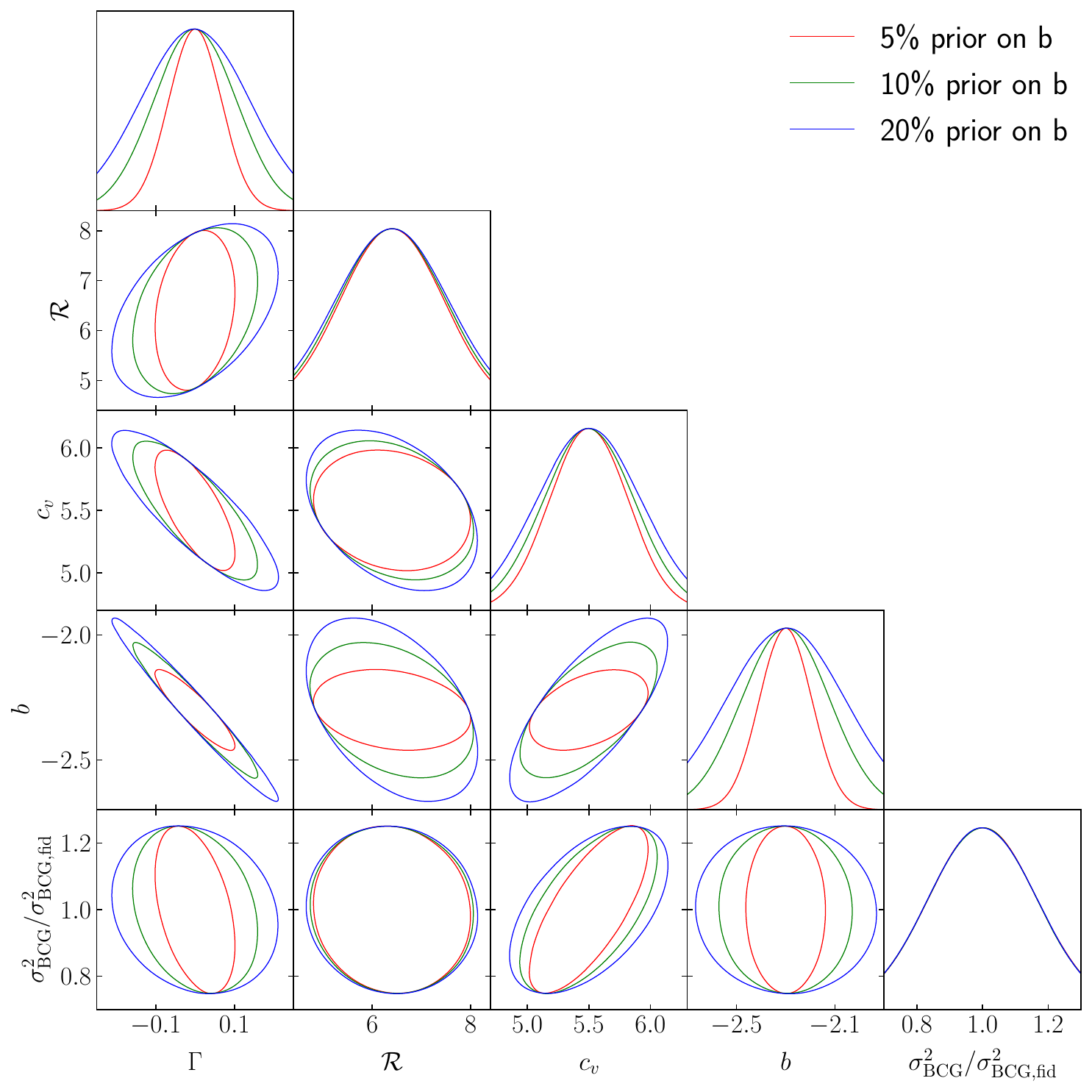}
    \caption{Joint 1$\sigma$ constraints on the full set of parameters. We assume a $20\%$ Gaussian prior on $\mathcal{R}$ and ${\sigma_{\rm BCG}^2}$, and three different choices of priors on $b$: $5\%$ (red), $10\%$ (green) and $20\%$ (blue). This plot is generated using the GetDist code~\cite{Lewis:2019xzd}.}
    \label{fig:triangle_plot}
\end{figure}

\end{appendices}

\newpage
\bibliography{grav_red_biblio_MG}

\end{document}